\documentclass[aps,pra,reprint,twocolumn,showpacs,floatfix,superscriptaddress]{revtex4-1}
\usepackage{amssymb,amsmath,amstext}                
\usepackage{graphicx}                                               
\usepackage{epstopdf}                                               
\usepackage[english]{babel}
\usepackage{color}                                                     
\usepackage{bm}                                                        
\usepackage{appendix}                                              
\usepackage[utf8]{inputenc}
\usepackage{bbold}
\usepackage{bbm}

\usepackage{latexsym}
\usepackage{xcolor}
\usepackage{braket}
\usepackage{subfigure}
\definecolor{lblue} {RGB}{51,71,158}

\usepackage[colorlinks=true,citecolor=blue,linkcolor=blue,urlcolor=lblue]{hyperref}

\newcommand{\be}{\begin{equation}}
\newcommand{\ee}{\end{equation}}

\begin{document}

\title{On intermediate statistics across many-body localization transition}

\author{Bitan De}
\affiliation{Institute of Theoretical Physics, Jagiellonian University in Krak\'ow,  \L{}ojasiewicza 11, 30-348 Krak\'ow, Poland }
\author{Piotr Sierant}
\affiliation{The Abdus Salam International Center for Theoretical Physics, Strada Costiera 11, 34151, Trieste, Italy}
\affiliation{Institute of Theoretical Physics, Jagiellonian University in Krak\'ow,  \L{}ojasiewicza 11, 30-348 Krak\'ow, Poland }
\author{Jakub Zakrzewski}
\email{jakub.zakrzewski@uj.edu.pl}
\affiliation{Institute of Theoretical Physics, Jagiellonian University in Krak\'ow,  \L{}ojasiewicza 11, 30-348 Krak\'ow, Poland }
\affiliation{Mark Kac Complex
Systems Research Center, Jagiellonian University in Krakow, Krak\'ow,
Poland. }

\date{\today}

                              
\begin{abstract}
The level statistics in the transition between delocalized and localized {phases of} many body interacting systems is {considered}. We recall the joint probability distribution for eigenvalues resulting from the statistical mechanics for energy level dynamics as introduced by Pechukas and Yukawa. The resulting single parameter analytic distribution is probed numerically {via Monte Carlo method}.  The resulting higher order spacing ratios are compared with data coming from different {quantum many body systems}. It is found that this Pechukas-Yukawa distribution compares favorably with {$\beta$--Gaussian ensemble -- a single parameter model of level statistics proposed recently in the context of disordered many-body systems.} {Moreover, the Pechukas-Yukawa distribution is also} only slightly inferior to the two-parameter $\beta$-h ansatz shown {earlier} to reproduce {level statistics of} physical systems remarkably well. 
\end{abstract}
\maketitle

\section{Introduction}
\label{intro}

Fritz Haake made fundamental contributions first to quantum optics (see e.g. \cite{Bonifacio71,Haake79}) then to the fast developing in the eighties of the last century area of Quantum Chaos not only with his seminal monograph on the subject \cite{Haakebook} but also with many original works from introducing a celebrated kicked top model \cite{Haake87} to providing a link between Gutzwiller's periodic orbit theory \cite{Gutzwiller71} and random matrix statistics \cite{Mueller04,Mueller05,Heusler07}.

Those late works provided a highlight of Haake's fascination of the link between spectral properties of physical models and random matrix theory. Of particular interest for him has always been level dynamics (see Chapter.6 of \cite{Haakebook}). The dependence of the energy levels of a Hamiltonian
$H(\lambda)$ 
on some parameter $\lambda$ may be viewed as the motion of interacting fictitious particles (levels) with ``time''  $\lambda$ following the original formulation of Pechukas \cite{Pechukas83} and Yukawa \cite{Yukawa85}.  For large matrices of {size $N \times N$}, in $N\rightarrow\infty$ limit, statistical mechanics is applicable to those fictitious particles. Haake devoted particular interest to   periodically driven (Floquet) systems also in this context,  in particular {since} a transition between level clustering for integrable systems to level repulsion (for quantally chaotic system) may be viewed  as a relaxation toward equilibrium \cite{Haake88}. 

For this transition, following the original Wigner $2\times 2$ matrices approach, Lenz and Haake found an interpolating spacing distribution \cite{Lenz91} which compared well with random matrix simulations adding a significant contribution to the topic which originated with the early work of  Rosenzweig and Porter \cite{Rosenzweig60}. The Rosenzweig-Porter distribution was an important twist on random matrices.
As it is well known \cite{Mehtabook,Haakebook, Stockmann} for generalized time reversal invariant systems (the case we shall solely concentrate on) the Gaussian orthogonal ensemble of random matrices (GOE) may be constructed {by} considering matrices with independent Gaussian distributed entries with the variance of the diagonal elements being twice the variance of the off-diagonal elements. GOE ensemble faithfully represent statistical properties of quantally chaotic system as conjectured by Bohigas, Giannoni, and Schmidt \cite{Bohigas84}.
The Rosenzweig-Porter ensemble interpolates between GOE and Poisson cases using a single parameter $\chi=\sigma^2/N$, where
$\sigma^2$ is the variance of off-diagonal elements and $N$ the matrix size  \cite{Rosenzweig60}. Lenz and Haake solution \cite{Lenz91} gives the nearest spacing distribution for such a $2\times 2$ matrices ensemble.

This is by no means a unique solution for the transition  between GOE and Poisson level statistics. The different models that were proposed over the years are reviewed in Section \ref{histo}. 
The renewed interest in the problem comes from a realization  that the transition between spectral properties of  ergodic many-body interacting systems and the so called many-body localized phase {\cite{Nandkishore15, Alet18, Abanin19} } may be described in the statistical sense by {an ensemble} interpolating between GOE and Poissonian case as in the case of chaotic systems or single particle Anderson localization.  In Section~\ref{statrev} we review the statistical approach of Yukawa \cite{Yukawa85} {following the} excellent textbook of H.-J. Stockman \cite{Stockmann}.
Such a Pechukas-Yukawa statistical mechanics model  provides another, {to the best of our knowledge not} tested yet, proposition for the interpolating ensemble. This approach has certain aesthetic advantages over other propositions opening also possibilities of
further studies going beyond the eigenvalue statistics. A critical comparison of this ensemble predictions with numerical data obtained for disordered Heisenberg chain and further examples for other systems are the content of Section \ref{numer}. We conclude giving future perspectives of the presented approach in Section \ref{conclu}.

\section{The interpolating ensembles and formulae} 
\label{histo}

The early approaches to level statistics often concentrated, as the work of Lenz and Haake \cite{Lenz91}, on the nearest neighbor level spacing
distribution. The distribution of normalized spacings $s_i=e_{i+1}-e_i$ where $e_i$ are the Hamiltonian matrix eigenvalues renormalized (in the process called the level unfolding \cite{Haakebook,Stockmann}) in such a way that their mean density is unity. Some proposed an {\it ad hoc} expressions as a popular Brody distribution \cite{Brody73} {which} surprisingly well fitted low resolution experimental data -- see e.g. \cite{Stockmann}. {In contrast,}  Berry and Robnik \cite{Berry84b} proposed {a} distribution that was based on a clear physical assumption of the separation between
``chaotic'' wavefunctions faithful to GOE and those localized on the regular part{s} of the phase space. This simple foundation resulted in the distribution which {was} shown to work well for mixed phase space situations in the deep semiclassical limit \cite{Prosen98}. Importantly, long range
correlations of energy levels can also be addressed within this model \cite{Prosen99}. 

{Over the years,} the most common approach {to construct an ensemble interpolating between GOE and uncorrelated Poisson level statistics} has been to modify {the} GOE distribution resigning from its beautiful property of being invariant under orthogonal transformations. The original Rozenzweig-Porter proposition \cite{Rosenzweig60}
is just one of them. In the original version a single parameter $\lambda=\sigma^2/N$ defined the ensemble of matrices of rank $N$ where 
$\sigma^2$ was the variance of off-diagonal matrix elements of the Hamiltonian matrix. For $\lambda\rightarrow 0$ one recovers the Poisson ensemble while for $\lambda=1$ the GOE {ensemble is reproduced} provided {that} the variance of the diagonal elements is equal to 2. 
{ Recently, modifications of the Rosenzweig-Porter ensemble allowing for a broad, log-normal distribution of the off-diagonal matrix elements,  have been shown to be valid for various types of many-body and Anderson localization transitions \cite{Khaymovich20, Kravtsov20, Khaymovich21}. }

A second proposition goes back to Seligman and coworkers \cite{Seligman84} who postulated that the variance of off-diagonal elements $a_{ij}$ should scale as $\exp\left[ - (i -j)^2/\sigma^2\right]$. For $\sigma\rightarrow 0 $ one recovers the Poisson case while {for} $\sigma\rightarrow \infty$ {the matrix belongs to}  GOE.

Another well-known approach is that of Guhr  \cite{Guhr96} who used supersymmetric techniques to express the two-level correlation function in the Poisson-GOE ensemble in terms of a double integral. In the meantime Casati and coworkers \cite{Casati90,Casati91} as well as Fyodorov and Mirlin \cite{fyodorov91}                         considered banded Gaussian random matrices as a useful tool in describing the transition, the corresponding parameter was $y=b^2/N$
with $b$ being the matrix bandwidth and $N$ its rank.

For a long time, the nearest neighbor spacings were treated as essential entities to capture the GOE-to-Poisson transition in the context of quantum chaos. An another relevant parameter is the number-variance, defined as {$\Sigma^2(L)=\langle n(L)^2\rangle-\langle n(L)\rangle^2$}, where $n(L)$ is the number of {(unfolded)} eigenvalues in an interval {of length} $L$. It was demonstrated that for a large system size $L$, {the number variance} scales as {$\chi L$}, where $\chi$ varies from $0$ to $1$ {during the}  Possion-to-GOE transitions \cite{Bogomolny01,Bogomolny11,Garcia16}. However, the unfolding process, implemented to compute $e_i$ {and the number variance}, {is based on a} separation of the density of states  $\rho(E)$ into {smooth} and fluctuating parts. The distinction between these two parts of $\rho(E)$ is to some extent arbitrary and may  lead to dubious outcomes see, for instance, {the differences between} local unfolding and Gaussian broadening  \cite{Gomez02}. Over the past years, there is a constant effort to tackle such problems 
 from the mathematical and computational viewpoints \cite{Irving,Vargas}.\\ 

 To simplify the problem, Huse and Oganesyan \cite{Oganesyan07} introduced an important measure known as gap ratio, defined as $r_n=min[\delta_n,\delta_{n-1}]/max[\delta_n,\delta_{n-1}]$, where $\delta_n=E_{n+1}-E_{n}$ is the energy gap between the consecutive energy levels. The dimensionless {gap-ratio}  is independent of the local density-of-states, $\rho(E)$, and the problem of unfolding is avoided. Employing an exact diagonalization, they also established that a spin system with time reversal invariance undergoes a phase transition from ergodic (with GOE {level statistics}) to the many-body localised (MBL) phase (signified by the uncorrelated Possonian statistics {of energy levels}) as the magnitude of disorder increases {(see also \cite{Santos04})}. In a further endeavor, Atas et.al. \cite{Atas13} {proposed a}  Wigner-like surmise {for the gap ratio statistics}. While {the} level statistics {across} the ergodic-to-MBL transition was still studied with  standard level spacing {distributions} \cite{Santos10}, the gap ratio became {gradually} a more popular tool \cite{Pal10,Mondaini15,Luitz15}. 
 
 The quest for a proper level statistics model to describe {ergodic-to-}MBL transition speeded up following the  work of  Serbyn and Moore \cite{Serbyn16} who argued that the flow of {level} statistics {between the GOE and Poisson limits occurs} in two stages: (1) A power-law type interaction between the eigenvalues {\cite{Kravtsov95} of range that decreases with increase of disorder strength   and (2) an appearance of semi-poissonian level statistics between the energy levels as an offshoot of local interactions between the energy levels (see also \cite{Bogomolny99,Bogomolny01}).
 
 Among the latest developments a so called $\beta$-Gaussian($\beta-G$) model was introduced \cite{Buijsman18} {as a model to describe level statistics across ergodic-to-MBL transition.  The $\beta-G$ model is dependent on a single} parameter $\beta$ {that characterizes} the pairwise interaction of the energy levels taking all the energy pairs in consideration.  A real value of $\beta\in[0,1]$ allows one to describe the whole crossover between GOE and Poisson level statistics. For a comparison of the performance of different models see \cite{Sierant19b}.  In our recent work \cite{Sierant20} , we have proposed a modified 2-parameter model, namely, the $\beta-h$ model, where the interaction between the {energy} levels is limited to a $h$ {neighboring energy levels}. This $2$-parameter model was shown to be superior to its $\beta-G$ counterpart since it reproduced not only the nearest neighbor spacing ratio but also the higher order spacing ratios defined by  \cite{Chavda14,Tekur18}:
\begin{equation}
r_n^i=min\bigg[\frac{E_{i+2n}-E_{i+n}}{E_{i+n}-E_{i}}, \frac{E_{i+n}-E_{i}}{E_{i+2n}-E_{i+n}}\bigg],
\label{s2e1}
\end{equation}
where $E_{i}$ stands for the $i$-th energy level.  In the present work we compare predictions of these recent models with the so called {Pechukas-Yukawa} ($P-Y$) level-dynamics approach and verify its underlying efficiency to reproduce the level statistics in terms of {the higher order spacing ratios} \eqref{s2e1}. In the next section we remind the derivation of the joint-probability-distribution  of the $P-Y$ model (which may be found, e.g. in \cite{Stockmann}) to make the paper self contained while later we compare the statistics generated with different models in the transition between delocalized and MBL cases.

\section{Statistical approach to level dynamics}
\label{statrev}

Consider a Hamiltonian dependent on parameter $\lambda$: $H(\lambda)=H_0+\lambda V$ for arbitrary
$H_0$ and $V$ and the fate of its eigenvalues as $\lambda$ is varied.  Differentiating the eigenvalue equation 
\begin{align}
H(\lambda) |a(\lambda)\rangle=x(\lambda)_a|a(\lambda)\rangle,
\label{eigen}
\end{align}
with respect to $\lambda$ (above $x(\lambda)_a$ is the eigenvalue corresponding to eigenvector 
$|a(\lambda)\rangle$) and taking appropriate scalar products one immediately gets
\begin{align}
\frac{d}{d\lambda} x_a \equiv \dot{x}_a=\langle a|V|a\rangle \equiv V_{aa}.
\label{e3}
\end{align}
Defining $p_a \equiv \dot{x}_a$ and looking at its $\lambda$ derivative yields
\begin{align}
\dot{p}_{a}=2\sum_{b\neq a}\frac{V_{ab} V_{ba}}{x_a-x_b}= 2\sum_{b\neq a}\frac{|f_{ab}|^2}{(x_a-x_b)^3}
\label{e4}
\end{align}
where the second equality involves an additional definition $f_{ab}=V_{ab}(x_a-x_b)$.
Most interestingly, differentiating $f_{ab}$ over $\lambda$, the set of differential equations closes:
 \begin{equation}
\dot{f}_{ab}=\sum_{r\neq a,b}^{}f_{ar}f_{rb}\bigg[\frac{1}{(x_a-x_r)^2}-\frac{1}{(x_b-x_r)^2}\bigg].
\label{e5}
\end{equation}
The set of equations \eqref{e3}-\eqref{e5} is known as the {Pechukas-Yukawa} equations. Identifying $\lambda$ as a fictitious time, $x_a$'s and $p_a$'s can be interpreted as positions and momenta of fictitious particles moving under the force {decaying} as $1/x^3$ with the distance (assuming $f_{ab}$'s are constant). This highly nonlinear set of of equations is neverthless integrable (for a discussion see \cite{Haakebook,Stockmann}).

Clearly, for $\lambda$ large
the eigenvalues of $H=H_0+\lambda V$ will be $V$ dominated and the variation of eigenvalues (and eigenvectors) trivializes. For that reason Haake \cite{Haakebook} introduces a slightly different $\lambda$ dependence (equivalent for small $\lambda$):  $H(\lambda)=\sqrt{f}(H_0+\lambda V)$ with $f=(1+\lambda^2)^{-1}$ while we shall use the form adopted from \cite{Zakrzewski93,Zakrzewski93b,Stockmann}
$H=H_0\cos(\lambda)+V\sin(\lambda)$.

Then the equations above become slightly modified with 
\begin{align}
 \dot{x}_a=\langle a|\dot{H}|a\rangle = p_a
\label{e3p}
\end{align}
and 
\begin{align}
\dot{p}_{a}=- x_a + 2\sum_{b\neq a}\frac{|f_{ab}|^2}{(x_a-x_b)^3}
\label{e4p}
\end{align}
with $f_{ab}=\langle a | \dot H | b \rangle (x_a-x_b)$. Thus the additional harmonic force binds the fictitious particles preventing their escape to infinity.

Pechukas-Yukawa equations corresponding to integrable system call for an appropriate statistical mechanics description which
should be realized within the generalized Giggs ensemble \cite{Vidmar16}. This requires an identification of the complete set of nontrivial constants of motion being in convolution. Such a possible approach is a song of the future. Here we rather follow Yukawa approach 
and consider  the simplest integrals of motion. One of them will be the total energy of the system of interacting classical particles
\begin{equation}
E=\frac{1}{2}\sum_{n}^{}(p_n^2+x_n^2)+\frac{1}{2}\sum_{n,m}\frac{|f_{nm}|^2}{(x_n-x_m)^2}.
\end{equation}
and the other  $Q=\frac{1}{2}Tr(F^2)$ called a total angular momentum,  for a justification see \cite{Stockmann}. Let us note that those are the only two second order constants of the motion.

With this two constants  the phase-space density is given by the Gibbs ensemble as 
\begin{equation}
\rho=\frac{1}{Z}\exp(-\beta E-\gamma Q),
\end{equation}
which in the presence of the harmonic binding potential  reads explicitly
\begin{equation}
\begin{split}
\rho=\frac{1}{Z}\exp(-\beta\bigg[\frac{1}{2}\sum_{n}^{}(p_n^2+x_n^2)+\frac{1}{2}\sum_{n,m}\frac{|f_{nm}|^2}{(x_n-x_m)^2}\bigg]\\
-\gamma \frac{1}{2}\sum_{n,m}^{}\bigg|f_{nm}\bigg|^2)
\end{split}
\end{equation}
Next, we can compute the joint probability distribution (JPD) of eigenvalues \cite{Stockmann} by integrating out the variables $p_n$ and $f_{nm}$ from the phase space density obtaining
\begin{equation}
\begin{split}
P(x_1,x_2,.....,x_n)\sim\prod_{n<m}^{}\bigg|\frac{(x_n-x_m)^2}{1+\frac{\gamma}{\beta}(x_m-x_n)^2}\bigg|^{\nu/2}\\ exp\bigg(-\frac{\beta}{2}\sum_{n}^{}x_n^2\bigg),
\end{split}
\label{s3e18}
\end{equation}
{where $\nu=1,2,4$ corresponds to three possible ensembles of Dyson. All three cases appear due to different properties of (integrated over) $f_{nm}$
variables. We shall consider the generalized time reversal invariant case of $\nu=1$ only.}
From the expression above, one can easily reach the Possonian distribution by putting $\gamma/\beta \gg 1$.  The distribution becomes
proportional to
\begin{equation}
P(x_1,x_2,.....,x_n)\sim\exp\big(-\frac{\beta}{2}\sum_{n}^{}x_n^2\big),
\end{equation}
while in the opposite limit,  $0<\gamma/\beta <<1$, the distribution mimics the Gaussian ensemble. 

\begin{equation}
P(x_1,x_2,.....,x_n)\sim\prod_{n>m}^{}\bigg|x_n-x_m\bigg|^{\nu} \exp\big(-\frac{\beta}{2}\sum_{n}^{}x_n^2\big).
\end{equation}

To obtain an ensemble that interpolates between GOE and Poisson level statistics we fix $\beta=\nu=1$ and denote $\gamma/\beta=10^p$. The distribution \eqref{s3e18} becomes
\begin{equation}
\begin{split}
P(x_1,x_2,.....,x_n)\sim\prod_{n<m}^{}\bigg|\frac{(x_n-x_m)^2}{1+10^p (x_m-x_n)^2}\bigg|^{1/2}\\
exp\bigg(-\frac{1}{2}\sum_{n}^{}x_n^2\bigg)
\end{split},
\label{s3e21}
\end{equation}
where $p=\log_{10}\frac{\gamma}{\beta}$ is the single free parameter of the distribution. The first term in the \eqref{s3e21} signifies the pairwise interaction between the particles and the exponential term provides the harmonic binding of the eigenvalues. This single parameter
distribution, resulting from Pechukas-Yukawa statistical mechanics for time reversal invariant case, will be tested  against other existing propositions as well as numerical data from interacting many-body models.

\section{Comparison with numerical data}
\label{numer}

For the simulation purpose, the eigenvalues are drawn by sampling \eqref{s3e21} with Metropolis-Hastings algorithm {\cite{Metropolis53, Hastings70} }for $p\in[-10, +15]$ with an initial sample size $N=500$ distributed over a linear chain. This moderate system size is chosen for convenience and speed. We consider about 100 eigenvalues in the middle of the spectrum. Since we consider high order gap ratios too, this sample size may be insufficient for $r_n$ with $n$ large. We consider the issue of the system size in the next Section.
The obtained eigenvalue distributions interpolate between a faithful semi-circle histogram of the density-of-states in the GOE limit (when $p\rightarrow -\infty$) and a Gaussian distribution for $p$ large and positive.  In the following we extract the higher order spacing ratios from \eqref{s3e21} and fit different physical models results with that distribution.  For comparison we use  the single parameter $\beta$-Gaussian model \cite{Buijsman18} as well as the two-parameter $\beta-h$ model shown to yield accurate representation of many body data \cite{Sierant20}.

\subsection{Random Heisenberg chain}

\begin{center}
	\begin{figure}[htb!]
\includegraphics[width=0.96\linewidth]{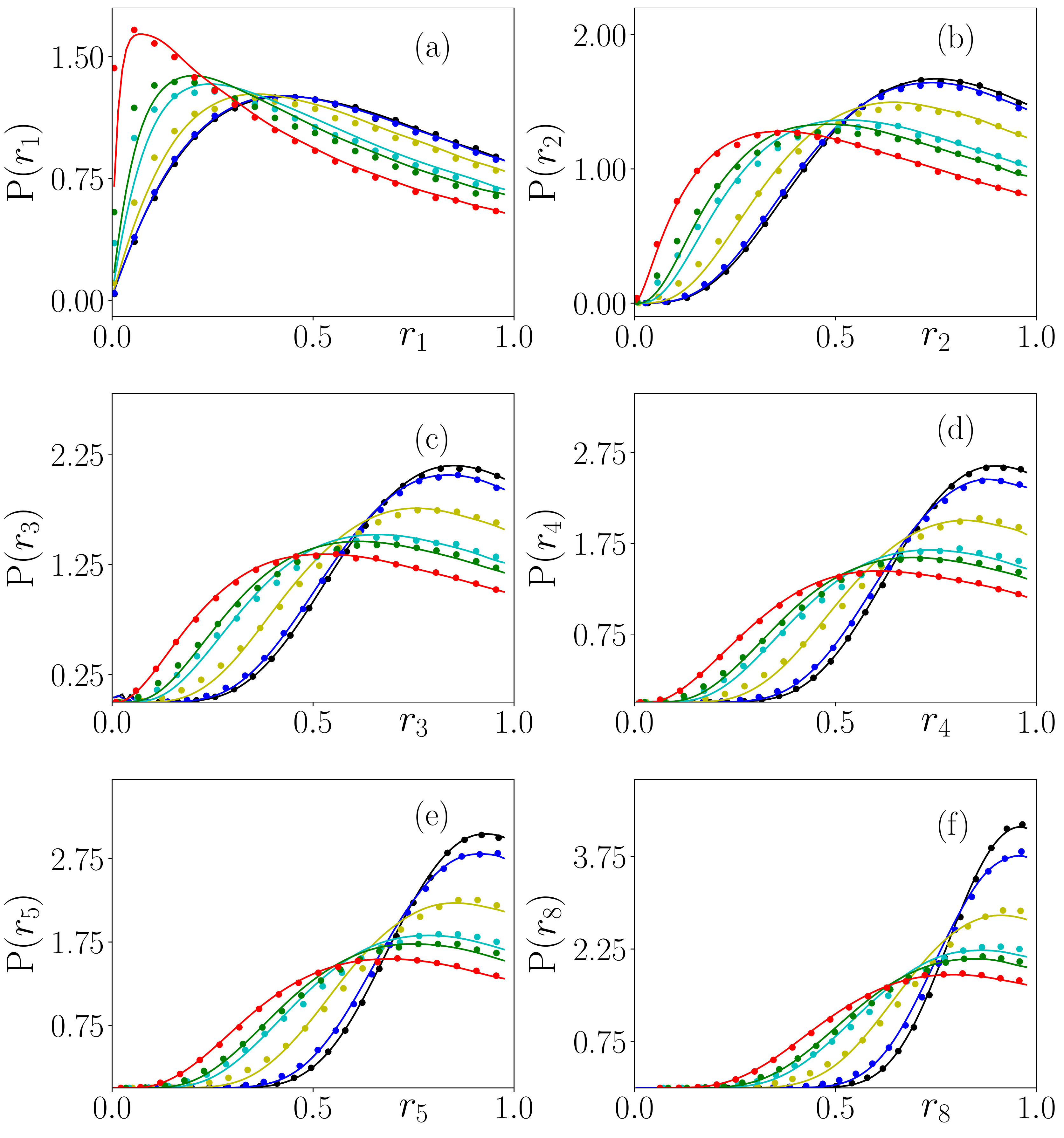}
		\caption{Probability distributions of the higher order spacing ratios, $r_n$, for $n=1,2,3,4,5,8$ {for}  a disordered Heisenberg spin chain with $L=18$ and the closest distributions given by $P-Y$ model. To facilitate the comparison the numerical data for the Heisenberg chain are shown by  bullets and the corresponding fits with the  $P-Y$ model by lines of the same color. {$W=1.8$ data (black bullets) are fitted with P-Y model with $p=2.3$; blue set is for $W=2.0$ ($p=3.1$); orange set is for $W=2.4$ ($p=4.4$); light green set for $W=2.8$ ($p=5.4$); dark green set for $W=3.0$ ($p=5.8$); and the red set for $W=4.0$ ($p=7.4$). }
		}
		\label{fig1}
	\end{figure}
\end{center}

As a first example we  take a disordered Heisenberg spin-$\frac{1}{2}$ chain, a paradigmatic model for MBL studies \cite{Luitz15,Luitz16,Serbyn16,Alet18,Mace19,Sierant19b}.  The Hamiltonian of the system reads 
\begin{equation}
\hat{H}_{H}=J\sum_{i=1}^{L}\vec{S}_{i}{\cdot}\vec{S}_{i+1}+\sum_{i=1}^{L}h_{i}{S}_{i}^{z}
\label{s4e1},
\end{equation}
where the first term represents the exchange interaction between the neighboring 1/2-spins with exchange coupling $J$ (normalized to unity). The last term constitutes the on-site disorder potential at site  $i$ with disorder $h_i$ drawn from uniform random distribution,  $h_i\in [-W,W]$. It is well established  that this system undergoes a transition from ergodic to MBL phase \cite{Luitz15,Luitz16,Serbyn16,Alet18,Sierant19b} for sufficiently large $W$. 

The eigenvalues of the Heisenberg chain are obtained for the  length $L=18$ with periodic boundary conditions (PBC) via the {shift-and-invert} method. In the due course, we use 2000 disorder realizations for each value of $W$. The gap ratios are calculated from approximately $500$ eigenvalues from the middle of the spectrum.  In Fig.~\ref{fig1}, we present a comparison between the  obtained gap-ratio distributions 
$P(r_n)$ and those resulting from the closest predictions coming from the $P-Y$ model \eqref{s3e21}. Different $n$ values allow us to explore correlations on different distances between eigenvalues. Let us stress that we use a single $p$ value and the resulting $P-Y$ distribution to fit all  $P(r_n)$ for $n\in[1,10]$ by minimizing the cumulative error between the gap-ratios of Heisenberg chain and the gap-ratios predicted by $P-Y$ model. 
As may be appreciated in Fig.~\ref{fig1}, the $P-Y$ distribution fits the data of the spin model remarkably well, in particular on the delocalized side of the transition (smaller $W$ values). Only close to the fully localized case some small deviations are seen.

\begin{center}
	\begin{figure}[htb!]
\includegraphics[width=0.96\linewidth]{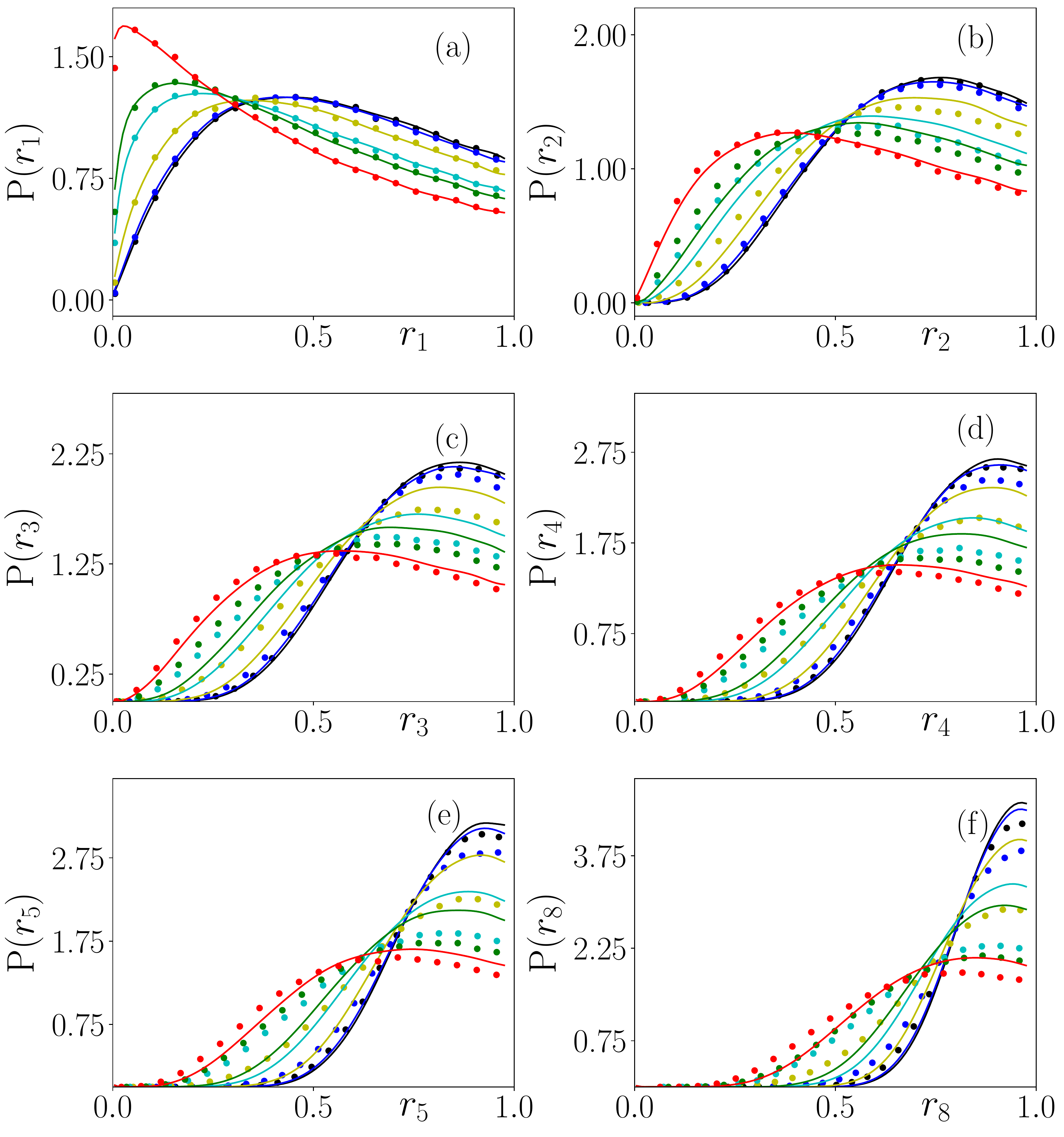}
		\caption{{Same as Fig.~\ref{fig1} but for the $\beta-G$ ensemble. The value of $'beta$ is fitted for $P(r_1)$ distribution as in \cite{Buijsman18}. Higher order gap ratios are clearly not reproduced satisfactorily. The numerical data for the Heisenberg chain are shown by  bullets and the  fits by  $\beta-G$ model by lines of the same color. $W=1.8$ data (black bullets) are fitted with $\beta-G$ model with $\beta=1.0$; blue set is for $W=2.0$ ($\beta=0.94$); orange set is for $W=2.4$ ($\beta=0.68$); light green set for $W=2.8$ ($\beta=0.38$); dark green set for $W=3.0$ ($\beta=0.26$); and the red set for $W=4.0$ ($\beta=0.06$).  }
		}
		\label{fignew1a}
	\end{figure}
\end{center}
To see how competitive $P-Y$ model is  we compare its predictions with those given by a single parameter  $\beta-G$ approach, in which
sampling of the distribution is much easier \cite{Buijsman18} as well as a two parameter family of $\beta-h$ distributions \cite{Sierant20}.
{For a most direct comparison with data in Fig.~\ref{fig1} we present similar figures for $\beta-G$ in  Fig.~\ref{fignew1a} and for $\beta-h$ ensemble in Fig.~\ref{fignew1b}. Even a casual inspection of Fig.~\ref{fignew1a} reveals that  $\beta-G$ ensemble fitted for $r_1$ (where it performs remarkably well) does not represent properly higher gap ratios. Clearly this ensemble does not reproducer short-range spacings and long-range correlations equally well. On the other hand $\beta-h$ 
ensemble performs at least as well as the P-Y distribution considered. }  

\begin{center}
	\begin{figure}[htb!]
\includegraphics[width=0.96\linewidth]{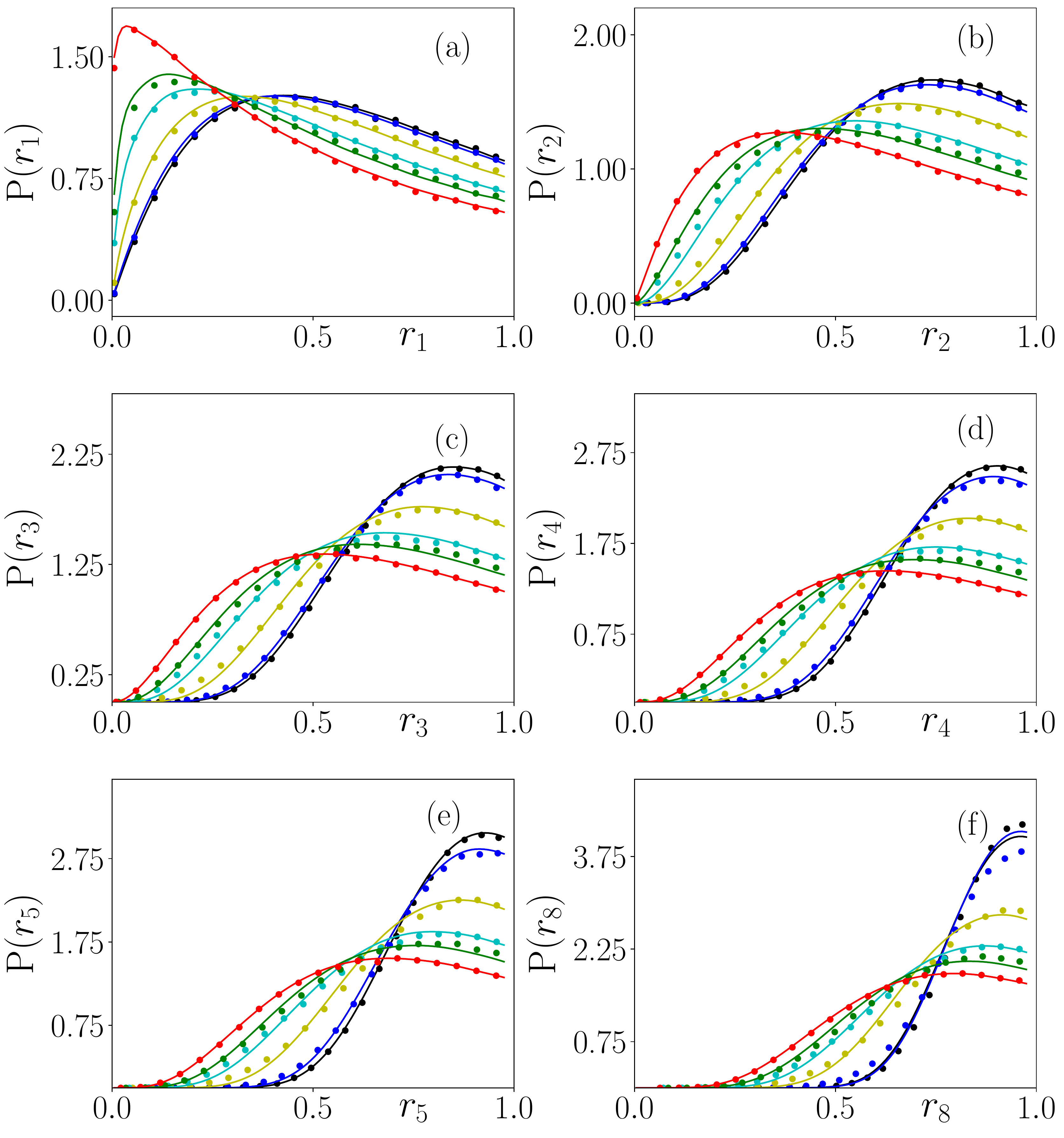}
		\caption{Same as Fig.~\ref{fig1} and Fig.~\ref{fignew1a} but for $\beta-h$ distridution which fits Heisenberg spin data remarkably well. {$W=1.8$ data (black bullets) are fitted with $\beta-h$ model with $\beta=1$, $h=10$; blue set is for $W=2.0$ ($\beta=0.96$, $h=6$); orange set is for $W=2.4$ ($\beta=0.74$, $h=2.7$); light green set for $W=2.8$ ($\beta=0.46$, $h=1.8$); dark green set for $W=3.0$ ($\beta=0.3$, $h=1.4$); and the red set for $W=4.0$ ($\beta=0.08$, $h=1.05$). Some of the data plotted were  presented in Fig.~3 of \cite{Sierant20} where $\beta-h$ model was introduced. }
		}
		\label{fignew1b}
	\end{figure}
\end{center}

\begin{figure}[htb!]
	\begin{center}
\includegraphics[width=0.96\linewidth]{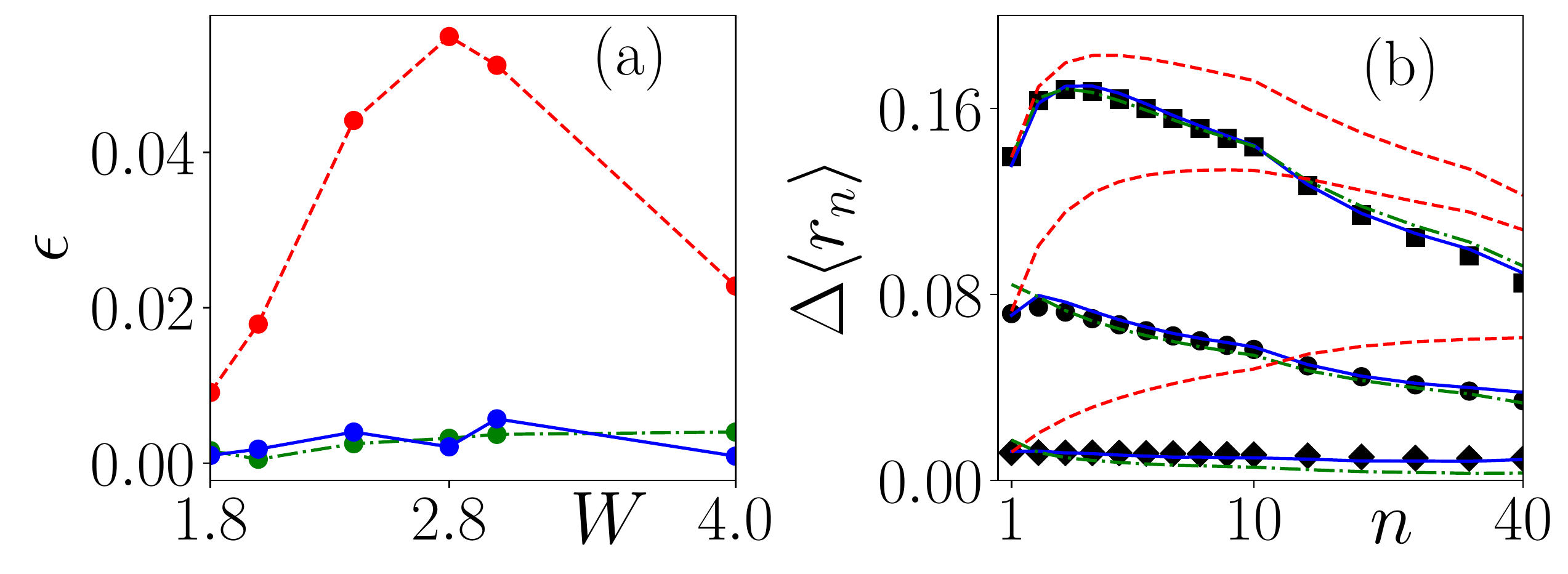}
	\end{center}
	\caption{(a) The  comparison of the cumulative error of the fit  $\epsilon$, \eqref{epsil} for all three studied distributions. Red dashed line corresponds to $\beta-G$ model, blue solid line for $\beta-h$ while green dashed-dot line for $P-Y$ model.  (b) The performance of different models compared for different mean gap ratios $\langle r_n\rangle$ as compared with the data for XXZ spin chain for $W=2.0$ (squares), $W=2.8$ (circles) and $W=4.0$ (diamonds) We plot the difference between the average $\langle r_n\rangle $ and the corresponding value for the Poisson distribution as it better represents the quality of fits. The lines correspond to different models consistently with panel (a).	}
	\label{fig2} 
\end{figure}

 For a  {global, quantitative}  comparison, we construct a measure of the fit quality, i.e., the average deviation 
 \be
 \epsilon \equiv \frac{1}{10}\sum_{n=1}^{10}\bigg[\langle r_n\rangle_M-\langle r_n\rangle_{H}\bigg],
 \label{epsil}
 \ee
 where $\langle r_n\rangle_{H}$ is the numerical average for the Heisenberg chain while $\langle r_n\rangle_M$ is the corresponding value coming from a fit to a given model, be it $P-Y$, $\beta -  h$ or $\beta-G$ version. {The obtained values of } $\epsilon$ {are} plotted 
 in Fig.~\ref{fig2}(a), 
 for $\beta-G$ model $\beta$ is fitted to reproduce as closely as possible $\langle r_1 \rangle_H$. Close to the delocalized GOE-like regime the resulting $\epsilon$ for that ensemble is comparable to both $P-Y$ and $\beta-h$ 
 values. Closer to the transition and on the localized side one may clearly observe that $\beta-G$ model performs poorly when compared with $P-Y$ and $\beta-h$ models. The latter is {only} slightly superior but it involves fitting of two parameters instead of one.

{Fig.~\ref{fig2}(b)} presents the average gap ratios $\langle r_n\rangle$ for $n\in[1,40]$ or rather, for a better visualization, a difference $\Delta \langle r_n \rangle =  \langle r_n\rangle - \langle r^{\rm P}_n\rangle$ where the latter corresponds to an analytic prediction for the mean gap ratio for a Poissonian ensemble \cite{Sierant20} (note that normalization factor is missing there).  On a first glance it is clear,  that the $\beta-G$ model is inaccurate beyond $r_1$ as {it severely overestimates  $\Delta \langle r_n \rangle$, which means that the long-range correlations between eigenvalues of  $\beta-G$ model are much stronger than the correlations of eigenvalues of the disordered Heisenberg spin chain }. Both $\beta-h$ and $P-Y$  reproduce higher {order} spacing ratios {much more accurately}.  {The} single parameter  $P-Y$ model is only slightly inferior to {the} $\beta-h$ {model}.

\begin{center}
	\begin{figure}[htb!]
\includegraphics[width=0.96\linewidth]{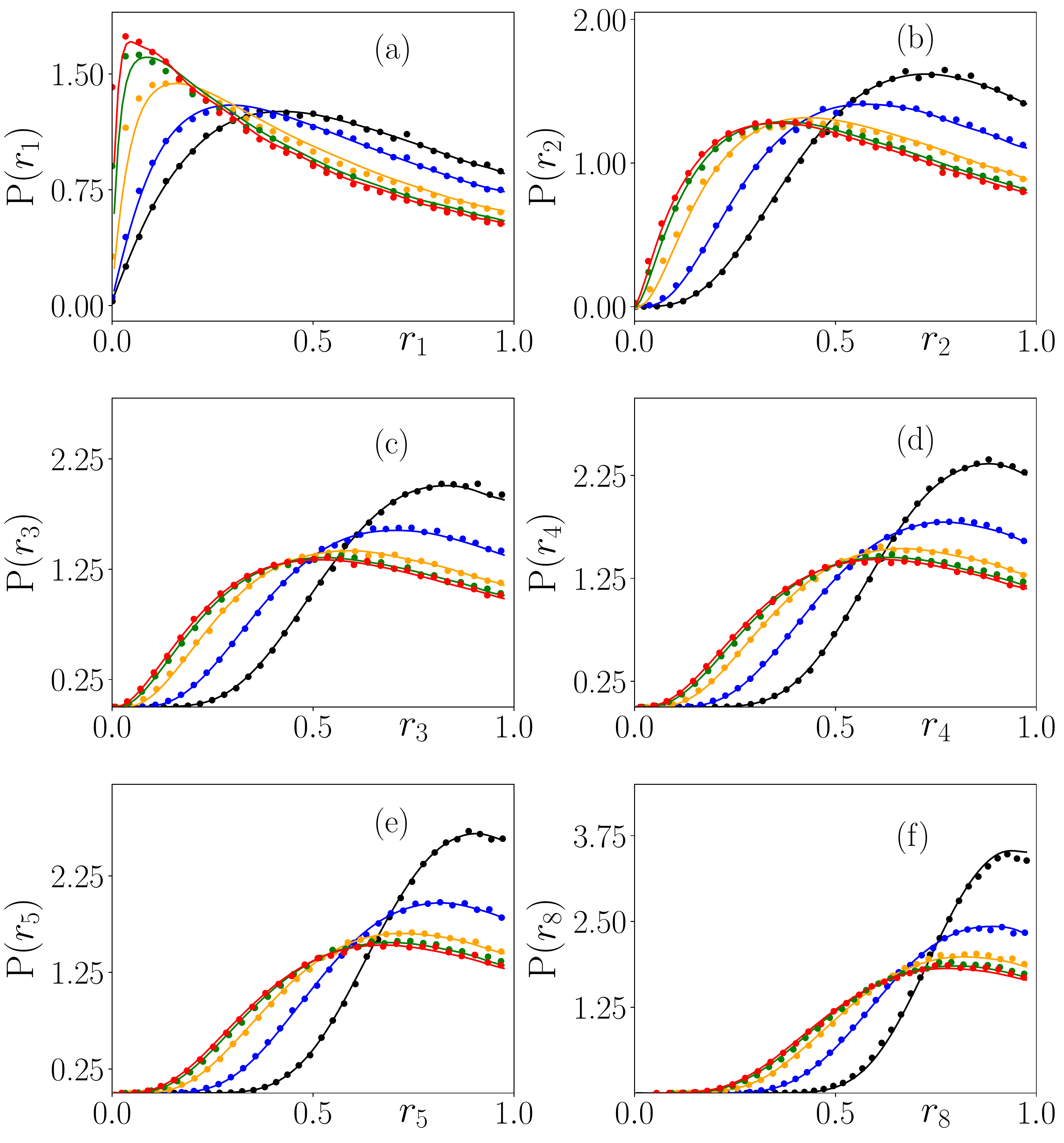}
		\caption{The distributions of the higher order gap-ratios of disordered quasi-periodic spin chain with $L=16$ and the best fit obtained in  the $P-Y$ model. The numerical data for the quasi-periodic chain for a given disorder $W$ are presented by  bullets and the resulting fits  by $P-Y$ model are shown by lines. $W=2.0$ data (black bullets) are fitted with $P-Y$ model with $p=3.4$; blue set is for $W=2.5$ ($p=5.0$); orange set is for $W=3.0$ ($p=6.2$);  green set for $W=3.5$ ($p=7.2$); and the red set for $W=4.0$ ($p=7.8$).  }
		\label{fig3}
	\end{figure}
\end{center}

\subsection{Quasi-periodic Heisenberg spin chain}

Consider the same Heisenberg chain with the Hamiltonian \eqref{s4e1} but now $h_i$ are not random but {taken} as
$h_{i}=(W/2)\cos(2\pi \chi i+\alpha)$, where  $i$ is the site index. Different realizations of the disorder correspond to different choices of $\alpha$ drawn from random uniform distribution on $[0,2\pi)$ interval. Contrary to the previous case, this quasi-periodic disorder is fully correlated. Still, for sufficiently large $W$ (dependent on the value of $\chi$ \cite{Guarrera07,Doggen19}) the system undergoes a transition to MBL phase. Such a quasiperiodic disorder was implemented in experiments of the Munich group \cite{Schreiber15,Luschen17} by placing an additional weak standing wave on top of the primary one forming the optical lattice. Then $\chi$ is the ratio of wavevectors of both laser wavelengths.
 
In  our study we take $\chi=(\sqrt{5}-1)/2$ and diagonalize the Hamiltonian for $L=16$ and open boundary conditions.  Like {in} the preceding case, we implement the shift-invert method to collect eigenvalues from $2000$ disorder realizations and the higher order spacing ratios are computed form $500$ eigenvalues taken from the middle of the spectrum. 
\begin{center}
	\begin{figure}[htb!]
	\includegraphics[width=0.96\linewidth]{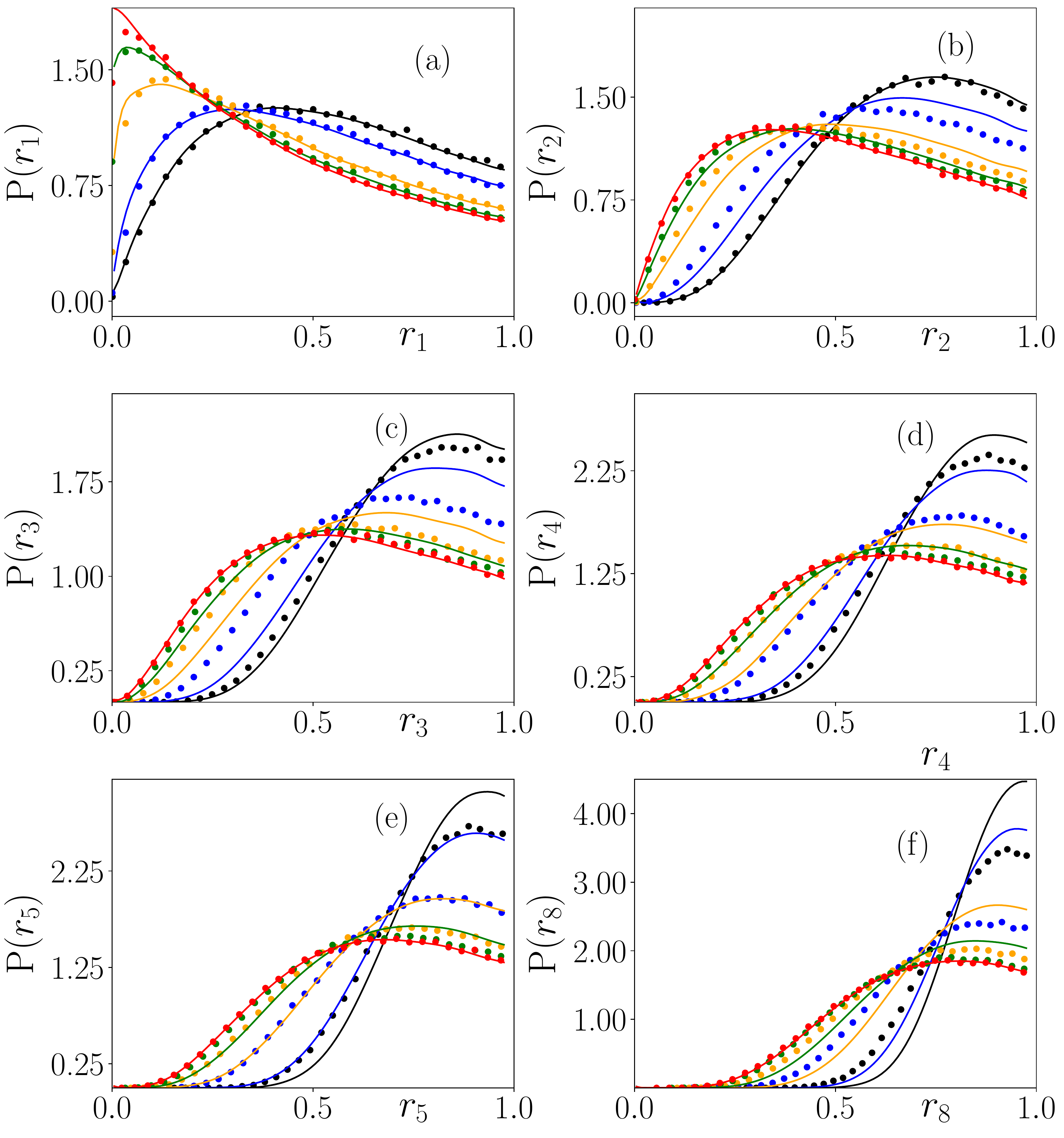}
		\caption{Same as Fig.~\ref{fig3} but a comparison of  the quasi-periodic chain data with their fits for $\beta-G$ ensemble is shown. As previously color differentiates between different disorder amplitudes $W$ (bullets) and different $beta-G$ distributions (lines) obtained by fitting $P(r_1)$.   $W=2.0$ data (black bullets) are fitted with $\beta-G$ model with $\beta=0.92$; blue set is for $W=2.5$ ($\beta=0.58$); orange set is for $W=3.0$ ($\beta=0.18$);  green set for $W=3.5$ ($\beta=0.06$); and the red set for $W=4.0$ ($\beta=0.01$).  }
		\label{fig3newa}
	\end{figure}
\end{center}

The whole analysis is performed {similarly to} the random disorder case.  We concentrate on the localized side of the crossover (for the system size studied) where the differences between {the considered} distributions show more clearly. {Fig.~\ref{fig3} shows the resulting fits for $P-Y$ ensemble showing that indeed it works very well also for this case.
For comparison results obtained for $\beta-G$ and $\beta-h$ ensembles are shown in Fig.~\ref{fig3newa} and Fig.~\ref{fig3newb}, respectively.}
The conclusions 
are very similar to the previous case  (although limited to smaller interval of $W$ taken). Firstly, {the single parameter} $P-Y$  model outperforms $\beta$-Gaussian proposition considerably, especially for higher order gap ratios. {Similarly to the uniform random disorder case,} $\beta$-Gaussian model works reasonably well only very close to the GOE limit (data corresponding to $W=1.5$ in Fig.~\ref{fig3}. Secondly, $P-Y$ distribution is clearly less effective than the phenomenological, two parameter $\beta-h$ approach in the whole interval of $W$ values. But, let us stress, this difference is quite small.
 
\begin{center}
	\begin{figure}[htb!]
\includegraphics[width=0.96\linewidth]{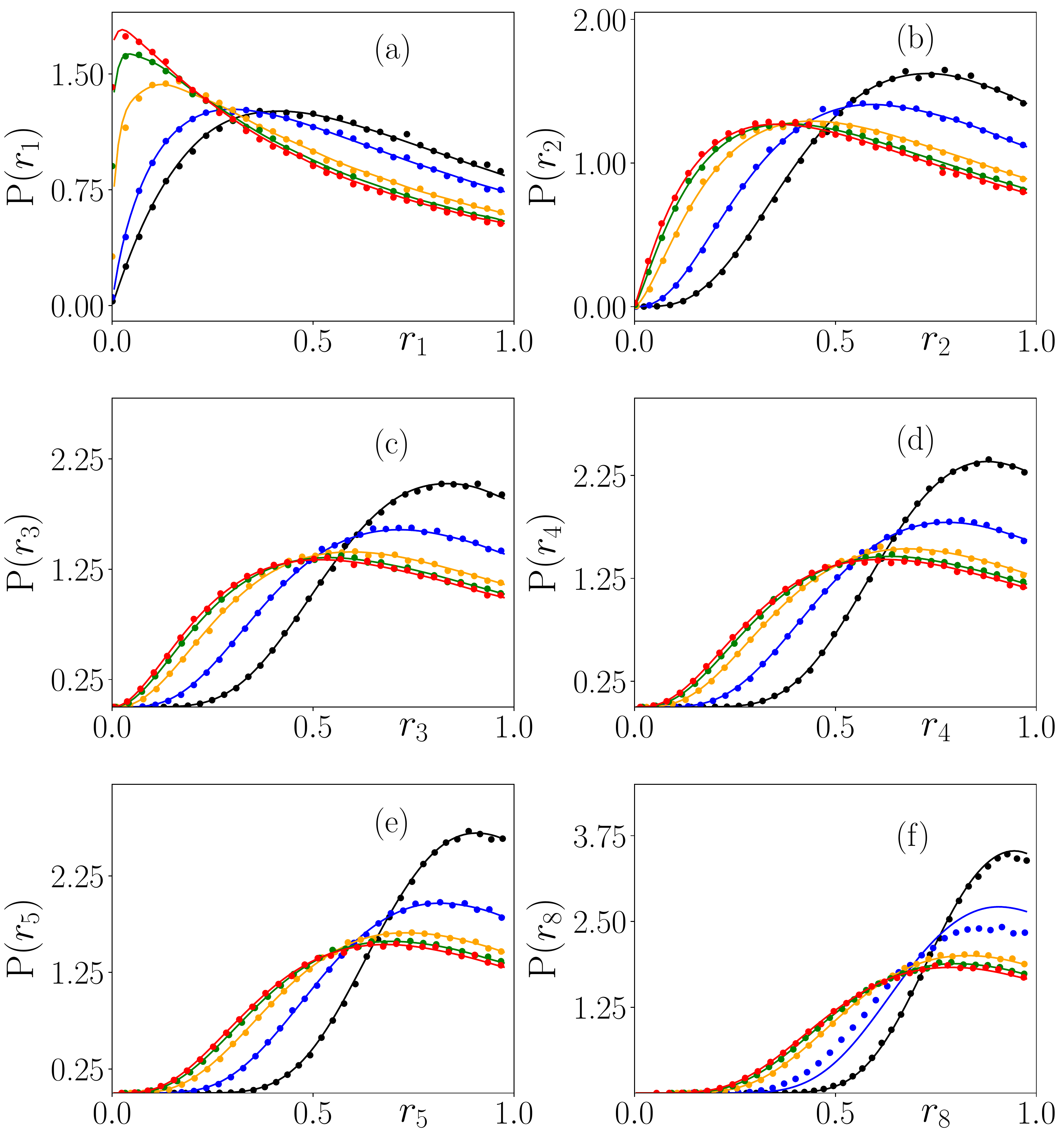}
		\caption{Same as Fig.~\ref{fig3} but a comparison of  the quasi-periodic chain data with their fits for $\beta-h$ model is presented. Different color corresponds to disorder amplitudes $W$ (bullets) and different  fits (lines) for the $\beta-h$ model.   $W=2.0$ data (black bullets) are fitted with $\beta-h$ model with $\beta=0.92$; blue set is for $W=2.5$ ($\beta=0.58$); orange set is for $W=3.0$ ($\beta=0.18$);  green set for $W=3.5$ ($\beta=0.06$); and the red set for $W=4.0$ ($\beta=0.01$).  }
		\label{fig3newb}
	\end{figure}
\end{center}

The global error comparison presented in Fig.~\ref{fig4} additionally confirms the above conclusions. 

It is worthwhile to comment more on the comparison of random and quasiperiodic cases. The crossover to MBL present in both models is quite well reproduced by $P-Y$ or $\beta-h$ statistical models despite the fact that {the systems with random and quasiperiodic disorder} behave quite differently on a microscopic level. This was observed in [64] inspecting the entanglement entropy statistics and further analysed{, on the level of level statistics,} in our earlier work [49]. We have shown there that
one may {distinguish the MBL transition in random and quasiperiodic disorder by examining the so called intersample variances which are} the sample-to-sample variances of gap ratios. Those show a peak in the crossover region for a random disorder case. This fact was attributed to the presence of rare Griffiths-type regions in systems with random disorder. The corresponding peak was absent in the quasiperiodic data supporting rare regions interpretation. One may pose a question{:} how {inter} sample variances of gap ratio behave 
in the statistical models such as $P-Y$ or $\beta-h$ models? We have verified that inter-sample variations of gap ratio practically do not change in the
GOE-Poisson transition for {the $P-Y$ model} by a direct evaluation. Our explanation of this  {discrepancy between behavior of a system with random disorder at the MBL transition and $P-Y$ model is the following.} {T}he rare regions affect variances {of the former}, {for which the} number of independent random {variables determining the Hamiltonian} is the same as {the} system size, $L$, very small compared to the Hilbert space {dimension}. For {the latter,} the number of random  {variables} in simulated matrices is {much larger and scales}  as a square of the matrix size making {the} sample-to-sample variations very small {across the whole GOE-Poisson crossover}.

\begin{figure}[htb!]
	\begin{center}
\includegraphics[width=0.96\linewidth]{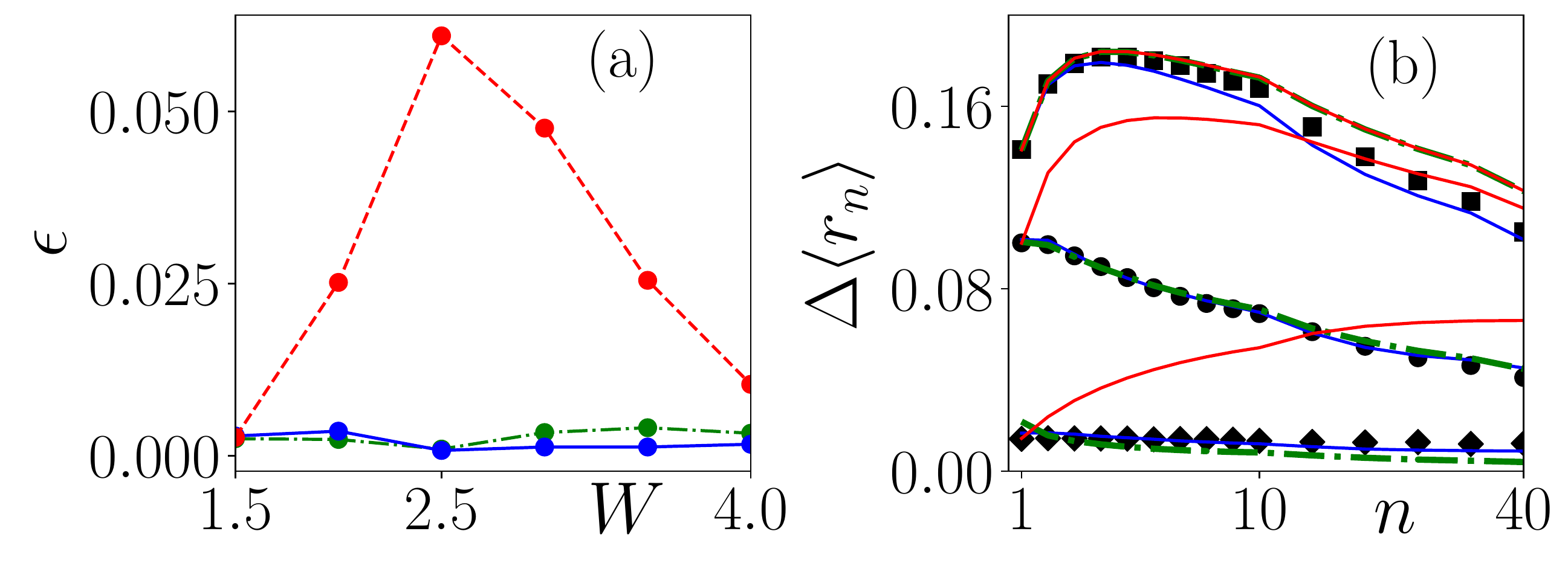}
	\end{center}
	\caption{Same as Fig.~\ref{fig2} but for quasiperiodic disorder. The cumulative error of $\beta-G$ distribution is much larger in the transition regime than for the other two distributions. Note that $P-Y$ again performs comparably to $\beta-h$ model. }
	\label{fig4}
\end{figure}


\subsection{Bose-Hubbard model}
\begin{figure}[htb!]
	\begin{center}
\includegraphics[width=0.96\linewidth]{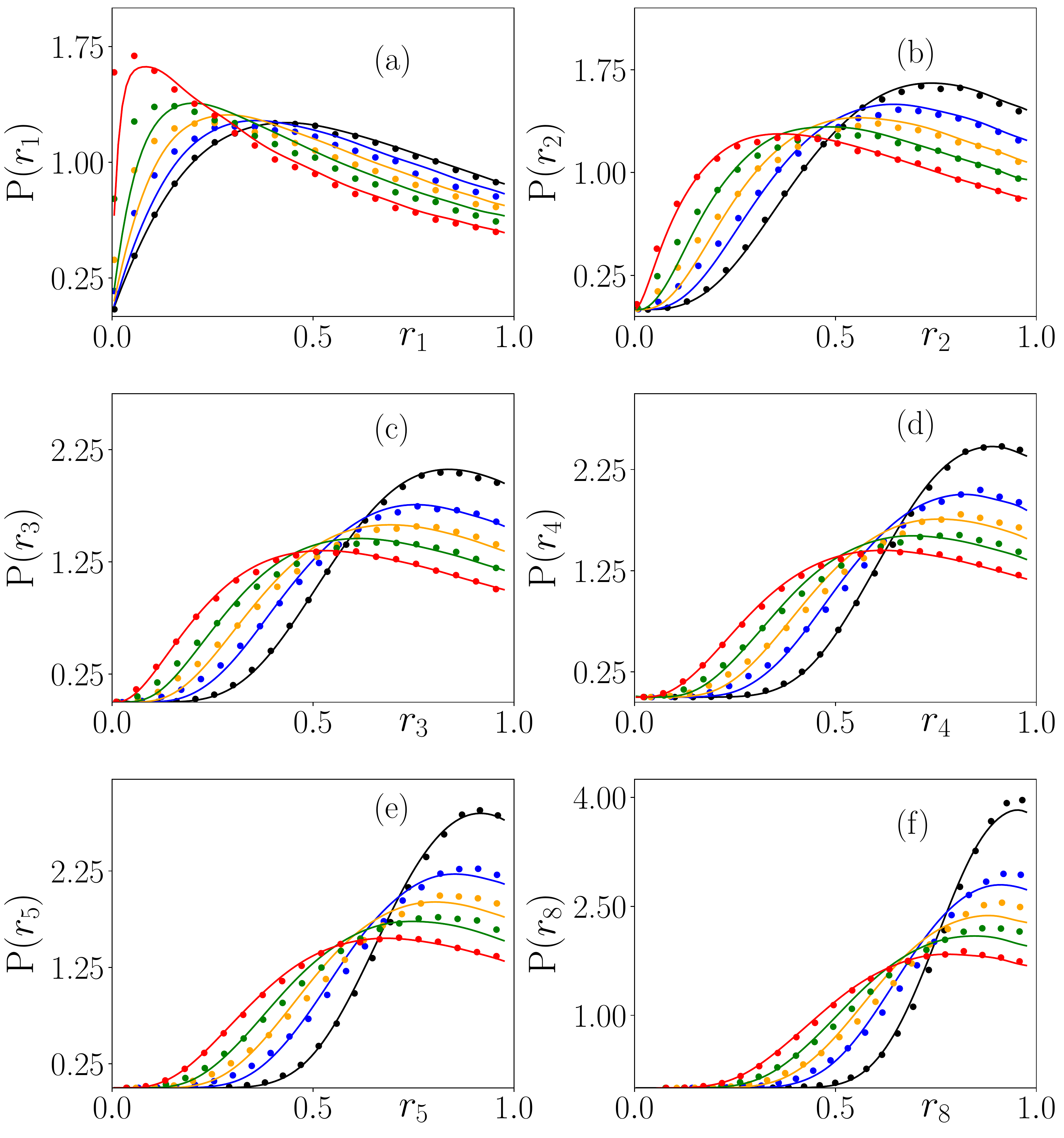}
	\end{center}
	\caption{$P(r_n)$ for $n=1,2,3,4,5,8$ for a Bose-Hubbard model with $L=8$ as fitted by the $P-Y$ model. The numerical data for BH model are shown by dots and the  fits of $P-Y$ model by lines. $W=7$ data (black bullets) are fitted with $P-Y$ model with $p=3.0$; blue set is for $W=10$ ($p=4.4$); orange set is for $W=12$ ($p=5.1$);  green set for $W=15$ ($p=5.8$); and the red set for $W=25$ ($p=7.3$). 
	}
	\label{fig5}
\end{figure}

As the last example we test $P-Y$ model on a different disordered system namely the Bose-Hubbard model with the Hamiltonian \cite{Sierant17,Sierant18,Orell19,Hopjan19}
\begin{equation}
\hat{H}_{BH}=-J\sum_{<i,j>}^{}\hat{a}_{i}^{\dagger}\hat{a}_{j}+\frac{U}{2}\sum_{i}^{}\hat{n}_{i}(\hat{n}_{i}-1)+\sum_{i}^{}\mu_{i}\hat{n}_{i},\label{s4e3}
\end{equation}
where $a_{k}^{\dagger}(a_{k})$ are  bosonic creation and annhilation operators at site $k$, $J$ is the tunneling and $U$ is the onsite interaction strength. We assume further on {$J=U=1$} while the chemical potential  $\mu_{i}$ is assumed to be uniformly randomly distributed 
within an interval $[-W,W].$  {As the considered previously spin chain} a transition from delocalized phase to MBL {is observed for such a disordered Bose-Hubbard model with increasing $W$ \cite{Sierant18}.}
\begin{figure}[htb!]
	\begin{center}
\includegraphics[width=0.96\linewidth]{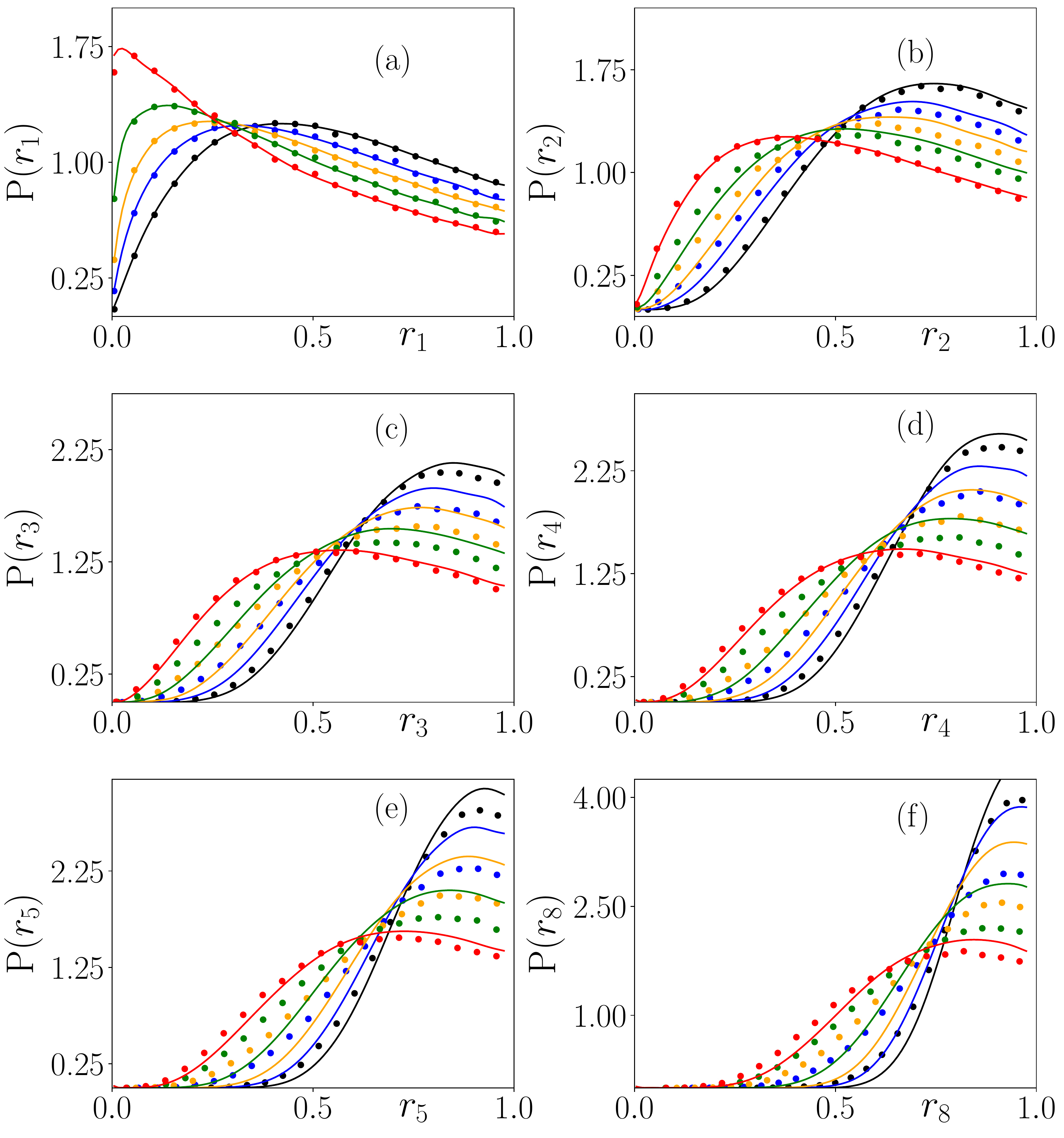}
	\end{center}
	\caption{Same as Fig.~\ref{fig5} but a comparison of  the Bose-Hubbard model data with their fits for $\beta-G$ ensemble is presented. Colors differentiate between different disorder amplitudes $W$ (bullets) and different $beta-G$ distributions (lines) obtained by fitting $P(r_1)$.   $W=7$ data (black bullets) are fitted with $\beta-G$ model with $\beta=0.94$; blue set is for $W=10$ ($\beta=0.62$); orange set is for $W=12$ ($\beta=0.42$);  green set for $W=15$ ($\beta=0.22$); and the red set for $W=25$ ($\beta=0.04$).  
}
	\label{fig5newa}
\end{figure}
\begin{figure}[htb!]
	\begin{center}
\includegraphics[width=0.96\linewidth]{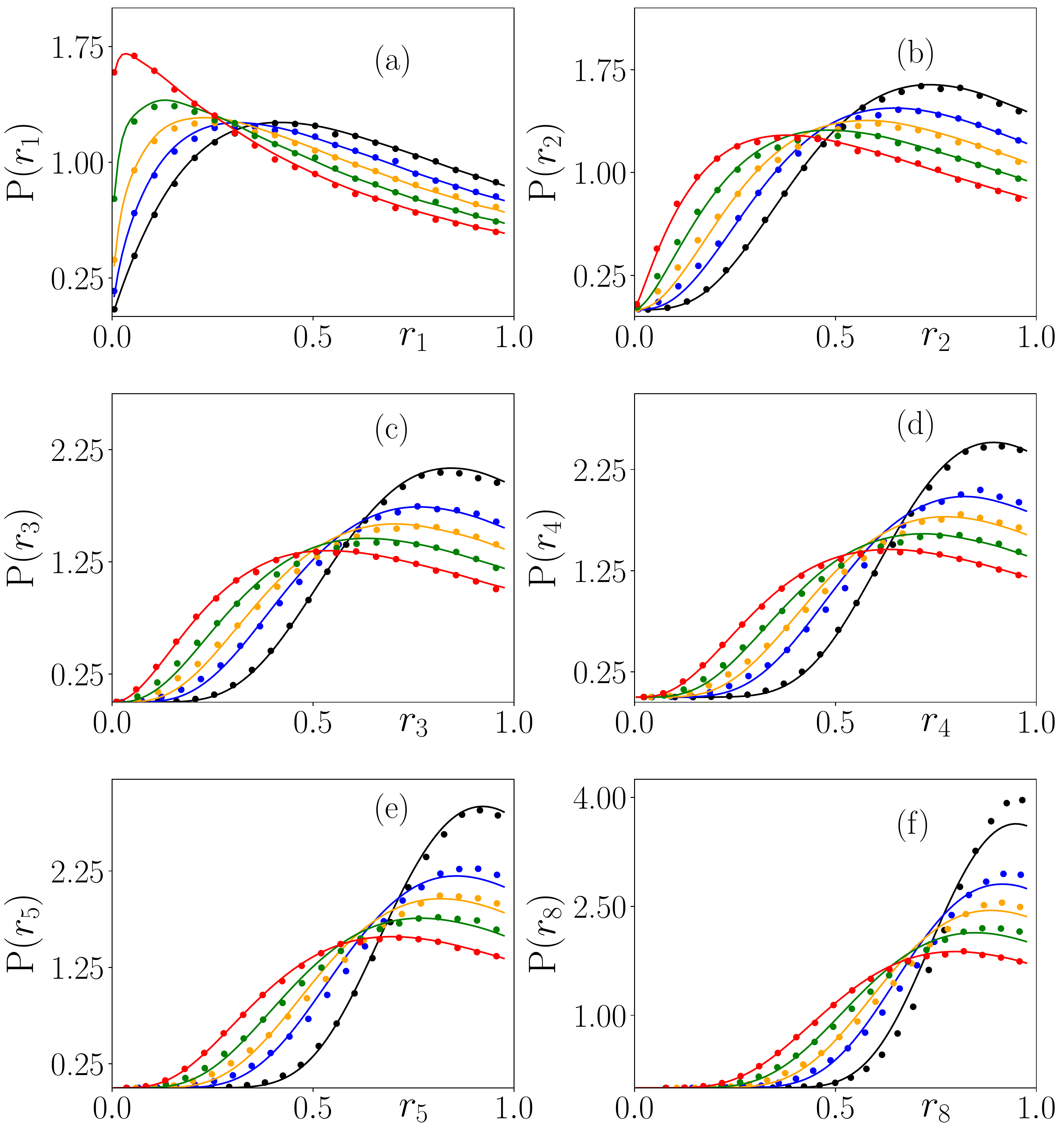}
	\end{center}
	\caption{{Similar comparison to Fig.~\ref{fig5newa} but fitting Bose-Hubbard data with the $\beta-h$ ensemble.
	$W=7$ data (black bullets) are fitted with $\beta=0.98$ and $h=6.5$; blue set is for $W=10$ ($\beta=0.70$, $h=2.6$); orange set is for $W=12$ ($\beta=0.46$, $h=2.4$);  green set for $W=15$ ($\beta=0.24$, $h=2.4$); and the red set for $W=25$ ($\beta=0.06$, $h=2$). 
	 Some of the data plotted were  presented in Fig.~8 of \cite{Sierant20} where $\beta-h$ model was introduced.
}
}
	\label{fig5newb}
\end{figure}

We consider a chain of $L=8$ sites at unit filling and eigenvalues are computed via exact-diagonalization method \cite{Sierant19b}. As previously, the gap ratios are determined from the eigenvalues collected from the middle of the spectrum for {no less than $500$} disorder realizations. The distribution of gap ratio $P(r_n)$ obtained are fitted with the $P-Y$ model as shown in Fig.~\ref{fig5} for $n=1,2,3,4,5,8$. {One may} observe a rather outstanding agreement between the data and the statistical {$P-Y$} model except for slight deviations close to full MBL (Poisson) cases. 
\begin{figure}[htb!]
	\begin{center}
\includegraphics[width=0.96\linewidth]{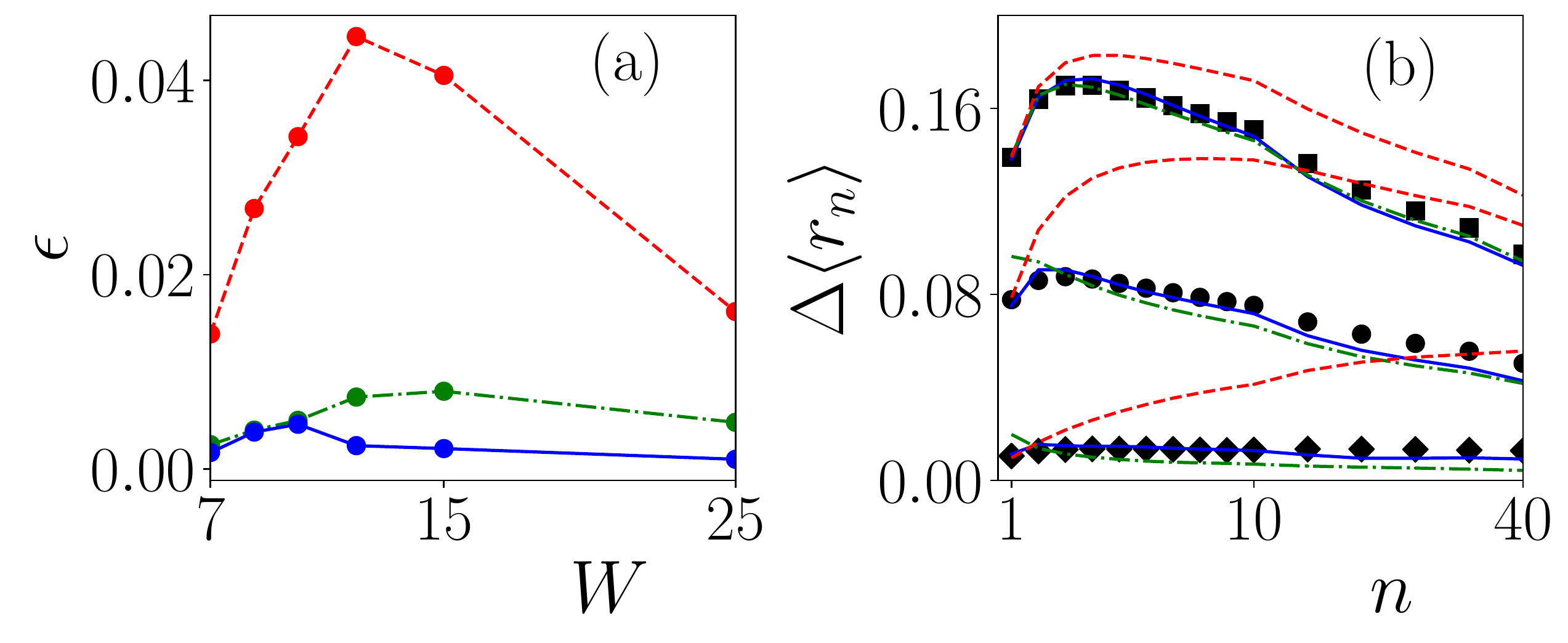}
	\end{center}
	\caption{Comparison of the error of all three distributions for Bose-Hubbard model. For that system $\beta-h$ model clearly outperforms $P-Y$ result, the latter is still acceptable as may be seen in the previous figure. $\beta-G$ model is performing much worse }
	\label{fig6}
\end{figure}

The corresponding comparison of numerical data for the disordered Bose-Hubbard model with $\beta-G$ and $\beta-h$ models are shown in Fig.~\ref{fig5newa} and Fig.~\ref{fig5newb}, respectively.
Similarly to {previously considered spin chains} , $P-Y$ model is clearly superior over the $\beta-G$ model in reproducing the data while being slightly outperformed by the two parameters $\beta-h$ model, as further summarised in Fig.~\ref{fig6}.

\section{Size considerations and the universal distribution}

The results presented for all three cases discussed show convincingly that the {single parameter} Pechukas-Yukawa distribution \eqref{s3e21} reproduces remarkably well distributions {of higher order spacing ratios} in the transition between GOE and Poisson regime. Let us recall the fact that we have consistently used data for $N=500$ ensemble. It is clear, from the form of the distribution \eqref{s3e21} that in the Poisson limit the density of states becomes a Gaussian with a unit variance independent of the system size. In the other, GOE limit the density of states follows a semicircle law in $(-\sqrt{2N},\sqrt{2N})$ interval being strongly $N$ dependent. In effect for any given set of data, the value of the fitted parameter $p$ in \eqref{s3e21} depends on $N$ chosen. 

This estetic drawback may be cured by rewriting \eqref{s3e21} as 

\begin{equation}
\begin{split}
P(x_1,x_2,.....,x_n)\sim\prod_{n<m}^{}\bigg|\frac{(x_n-x_m)^2}{1+10^Y (x_m-x_n)^2/\Delta^2}\bigg|^{1/2}\\
exp\bigg(-\frac{1}{2}\sum_{n}^{}x_n^2\bigg),
\end{split}
\label{s3e21a}
\end{equation}
where in the denominator the factor $10^p$ is replaced by $10^Y/\Delta^2$, where $\Delta$ is the mean spacing of $\{x_n\}$. This defines a new single parameter $Y$ characterizing the distribution, while $\Delta = \Delta(N,Y)$ can be easily obtained from generated sequence of $x_n$.  We have verified that such an approach makes the obtained distributions $N$ independent as visualised in Fig.~\ref{fignew}
which presents the average gap ratios $\langle r_i\rangle$ for $i=1,3$ for $N=500$ and $N=2000$.

\begin{figure}[htb!]
	\begin{center}
	\includegraphics[width=0.85\linewidth]{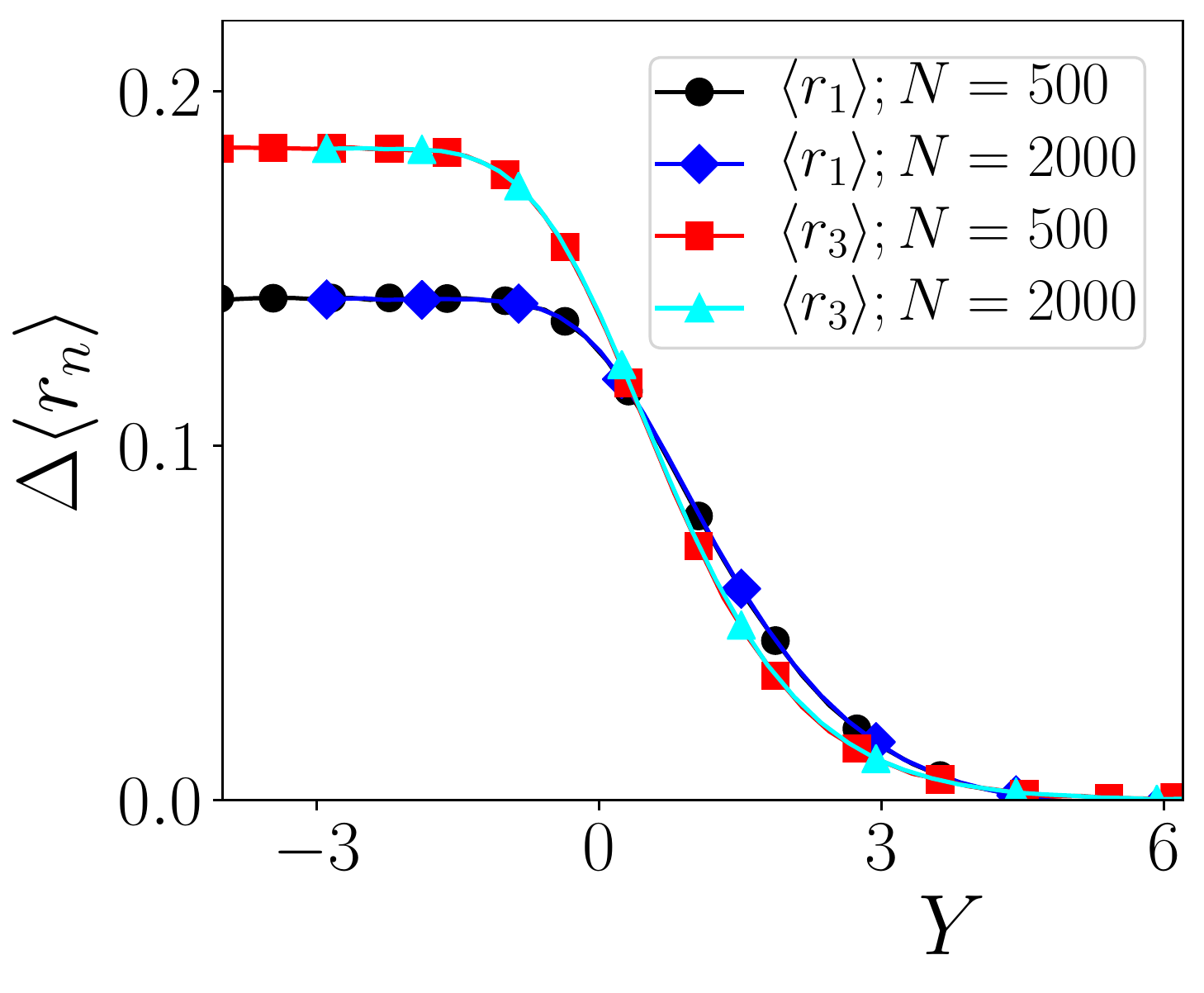}
	\end{center}
	\caption{The mean gap ratios {$\left \langle r_1 \right \rangle$ and $\left \langle r_3 \right \rangle $} offset by their Poisson value {$r_1^P\approx0.3862$} and $r_3^P\approx0.5625$ for $N=500$ and $N=2000$ obtained across the MBL transition using the universal distribution \eqref{s3e21a}.
	}
	\label{fignew}
\end{figure}

\section{Conclusions}
\label{conclu}

The aim of this work was to investigate how the distributions obtained from  JPD of eigenvalues resulting from standard Pechukas-Yukawa statistical approach compare with currently discussed models of level statistics interpolating between GOE and integrable cases. The nice feature of $P-Y$ model, as given by \eqref{s3e21}, is that a single parameter, denoted by $p$ allows for an interpolation in the whole interval between GOE and Poisson limits. The distribution studied results in statistics closely following the numerical data and in this respect is clearly superior to other single parameter model such as $\beta$-Gaussian proposition \cite{Buijsman18}. The latter has a clear advantage that probing it does not require the Metropolis algorithm, so probing it is relative simple. We also made a comparison with the so called $\beta-h$ model \cite{Sierant20} which quite accurately represents numerical data. The difficulty of applying the latter are comparable. It is quite rewarding to see that a single parameter $P-Y$ model yields results only slightly inferior to two-parameters $\beta-h$ approach.

It is worth stressing that while we concentrated on eigenvalue statistics, the Pechukas-Yukawa JPD contains also information on distribution of matrix elements $f_{nm}$ of a specific operator $F=[H,\dot H]$. The latter is manifestly time independent - thus matrix elements $f_{nm}$
carry information on eigenstates dependence on the parameter. Indeed Nakamura and Lakshmanan \cite{Nakamura86} {rephrased} the dynamics in terms of eigenvalues and components of eigenvectors. Thus JPD integrated over eigenvalues may yield prediction on eigenvector statistics in the transition regime (or alternatively on matrix elements of a generic operator).

Last but not least let us mention that one could aim at more ``proper'' analysis of integrable model given by \eqref{e3}-\eqref{e5} with the appropriate generalized Gibbs ensemble \cite{Vidmar16} provided a proper identification of all independent constants of the motion is made. 
A more modest approach will just include some of them going beyond the presented here quadratic terms approach.

The presented study may be generalized to cases with broken generalized time-reversal invariance as well as to an analogous approach possible for time-periodic Floquet systems.

\begin{acknowledgments}
J.Z. thanks Yan Fyodorov for suggestions on the literature of the subject and Mariusz Gajda and Krzysztof Sacha for discussions on Metropolis algorithm implementation.
The support of  PL-Grid Infrastructure is acknowledged.
This research has been supported by 
 National Science Centre (Poland) under project  2019/35/B/ST2/00034 (B.D., J.Z.). P.S. acknowledges the support of  Foundation  for
Polish   Science   (FNP)   through   scholarship   START.
 \end{acknowledgments}


\begin{thebibliography}{68}%
\makeatletter
\providecommand \@ifxundefined [1]{%
 \@ifx{#1\undefined}
}%
\providecommand \@ifnum [1]{%
 \ifnum #1\expandafter \@firstoftwo
 \else \expandafter \@secondoftwo
 \fi
}%
\providecommand \@ifx [1]{%
 \ifx #1\expandafter \@firstoftwo
 \else \expandafter \@secondoftwo
 \fi
}%
\providecommand \natexlab [1]{#1}%
\providecommand \enquote  [1]{``#1''}%
\providecommand \bibnamefont  [1]{#1}%
\providecommand \bibfnamefont [1]{#1}%
\providecommand \citenamefont [1]{#1}%
\providecommand \href@noop [0]{\@secondoftwo}%
\providecommand \href [0]{\begingroup \@sanitize@url \@href}%
\providecommand \@href[1]{\@@startlink{#1}\@@href}%
\providecommand \@@href[1]{\endgroup#1\@@endlink}%
\providecommand \@sanitize@url [0]{\catcode `\\12\catcode `\$12\catcode
  `\&12\catcode `\#12\catcode `\^12\catcode `\_12\catcode `\%12\relax}%
\providecommand \@@startlink[1]{}%
\providecommand \@@endlink[0]{}%
\providecommand \url  [0]{\begingroup\@sanitize@url \@url }%
\providecommand \@url [1]{\endgroup\@href {#1}{\urlprefix }}%
\providecommand \urlprefix  [0]{URL }%
\providecommand \Eprint [0]{\href }%
\providecommand \doibase [0]{http://dx.doi.org/}%
\providecommand \selectlanguage [0]{\@gobble}%
\providecommand \bibinfo  [0]{\@secondoftwo}%
\providecommand \bibfield  [0]{\@secondoftwo}%
\providecommand \translation [1]{[#1]}%
\providecommand \BibitemOpen [0]{}%
\providecommand \bibitemStop [0]{}%
\providecommand \bibitemNoStop [0]{.\EOS\space}%
\providecommand \EOS [0]{\spacefactor3000\relax}%
\providecommand \BibitemShut  [1]{\csname bibitem#1\endcsname}%
\let\auto@bib@innerbib\@empty
\bibitem [{\citenamefont {Bonifacio}\ \emph {et~al.}(1971)\citenamefont
  {Bonifacio}, \citenamefont {Schwendimann},\ and\ \citenamefont
  {Haake}}]{Bonifacio71}%
  \BibitemOpen
  \bibfield  {author} {\bibinfo {author} {\bibfnamefont {R.}~\bibnamefont
  {Bonifacio}}, \bibinfo {author} {\bibfnamefont {P.}~\bibnamefont
  {Schwendimann}}, \ and\ \bibinfo {author} {\bibfnamefont {F.}~\bibnamefont
  {Haake}},\ }\href {\doibase 10.1103/PhysRevA.4.302} {\bibfield  {journal}
  {\bibinfo  {journal} {Phys. Rev. A}\ }\textbf {\bibinfo {volume} {4}},\
  \bibinfo {pages} {302} (\bibinfo {year} {1971})}\BibitemShut {NoStop}%
\bibitem [{\citenamefont {Haake}\ \emph {et~al.}(1979)\citenamefont {Haake},
  \citenamefont {King}, \citenamefont {Schr\"oder}, \citenamefont {Haus},\ and\
  \citenamefont {Glauber}}]{Haake79}%
  \BibitemOpen
  \bibfield  {author} {\bibinfo {author} {\bibfnamefont {F.}~\bibnamefont
  {Haake}}, \bibinfo {author} {\bibfnamefont {H.}~\bibnamefont {King}},
  \bibinfo {author} {\bibfnamefont {G.}~\bibnamefont {Schr\"oder}}, \bibinfo
  {author} {\bibfnamefont {J.}~\bibnamefont {Haus}}, \ and\ \bibinfo {author}
  {\bibfnamefont {R.}~\bibnamefont {Glauber}},\ }\href {\doibase
  10.1103/PhysRevA.20.2047} {\bibfield  {journal} {\bibinfo  {journal} {Phys.
  Rev. A}\ }\textbf {\bibinfo {volume} {20}},\ \bibinfo {pages} {2047}
  (\bibinfo {year} {1979})}\BibitemShut {NoStop}%
\bibitem [{\citenamefont {Haake}(2010)}]{Haakebook}%
  \BibitemOpen
  \bibfield  {author} {\bibinfo {author} {\bibfnamefont {F.}~\bibnamefont
  {Haake}},\ }\href@noop {} {\emph {\bibinfo {title} {Quantum Signatures of
  Chaos}}}\ (\bibinfo  {publisher} {Springer, Berlin},\ \bibinfo {year}
  {2010})\BibitemShut {NoStop}%
\bibitem [{\citenamefont {Haake}\ \emph {et~al.}(1987)\citenamefont {Haake},
  \citenamefont {Ku{\'{s}}},\ and\ \citenamefont {Scharf}}]{Haake87}%
  \BibitemOpen
  \bibfield  {author} {\bibinfo {author} {\bibfnamefont {F.}~\bibnamefont
  {Haake}}, \bibinfo {author} {\bibfnamefont {M.}~\bibnamefont {Ku{\'{s}}}}, \
  and\ \bibinfo {author} {\bibfnamefont {R.}~\bibnamefont {Scharf}},\ }\href
  {\doibase 10.1007/BF01303727} {\bibfield  {journal} {\bibinfo  {journal}
  {Zeitschrift f{\"u}r Physik B Condensed Matter}\ }\textbf {\bibinfo {volume}
  {65}},\ \bibinfo {pages} {381} (\bibinfo {year} {1987})}\BibitemShut
  {NoStop}%
\bibitem [{\citenamefont {Gutzwiller}(1971)}]{Gutzwiller71}%
  \BibitemOpen
  \bibfield  {author} {\bibinfo {author} {\bibfnamefont {M.~C.}\ \bibnamefont
  {Gutzwiller}},\ }\href {\doibase 10.1063/1.1665596} {\bibfield  {journal}
  {\bibinfo  {journal} {Journal of Mathematical Physics}\ }\textbf {\bibinfo
  {volume} {12}},\ \bibinfo {pages} {343} (\bibinfo {year} {1971})}\BibitemShut
  {NoStop}%
\bibitem [{\citenamefont {M\"uller}\ \emph {et~al.}(2004)\citenamefont
  {M\"uller}, \citenamefont {Heusler}, \citenamefont {Braun}, \citenamefont
  {Haake},\ and\ \citenamefont {Altland}}]{Mueller04}%
  \BibitemOpen
  \bibfield  {author} {\bibinfo {author} {\bibfnamefont {S.}~\bibnamefont
  {M\"uller}}, \bibinfo {author} {\bibfnamefont {S.}~\bibnamefont {Heusler}},
  \bibinfo {author} {\bibfnamefont {P.}~\bibnamefont {Braun}}, \bibinfo
  {author} {\bibfnamefont {F.}~\bibnamefont {Haake}}, \ and\ \bibinfo {author}
  {\bibfnamefont {A.}~\bibnamefont {Altland}},\ }\href {\doibase
  10.1103/PhysRevLett.93.014103} {\bibfield  {journal} {\bibinfo  {journal}
  {Phys. Rev. Lett.}\ }\textbf {\bibinfo {volume} {93}},\ \bibinfo {pages}
  {014103} (\bibinfo {year} {2004})}\BibitemShut {NoStop}%
\bibitem [{\citenamefont {M\"uller}\ \emph {et~al.}(2005)\citenamefont
  {M\"uller}, \citenamefont {Heusler}, \citenamefont {Braun}, \citenamefont
  {Haake},\ and\ \citenamefont {Altland}}]{Mueller05}%
  \BibitemOpen
  \bibfield  {author} {\bibinfo {author} {\bibfnamefont {S.}~\bibnamefont
  {M\"uller}}, \bibinfo {author} {\bibfnamefont {S.}~\bibnamefont {Heusler}},
  \bibinfo {author} {\bibfnamefont {P.}~\bibnamefont {Braun}}, \bibinfo
  {author} {\bibfnamefont {F.}~\bibnamefont {Haake}}, \ and\ \bibinfo {author}
  {\bibfnamefont {A.}~\bibnamefont {Altland}},\ }\href {\doibase
  10.1103/PhysRevE.72.046207} {\bibfield  {journal} {\bibinfo  {journal} {Phys.
  Rev. E}\ }\textbf {\bibinfo {volume} {72}},\ \bibinfo {pages} {046207}
  (\bibinfo {year} {2005})}\BibitemShut {NoStop}%
\bibitem [{\citenamefont {Heusler}\ \emph {et~al.}(2007)\citenamefont
  {Heusler}, \citenamefont {M\"uller}, \citenamefont {Altland}, \citenamefont
  {Braun},\ and\ \citenamefont {Haake}}]{Heusler07}%
  \BibitemOpen
  \bibfield  {author} {\bibinfo {author} {\bibfnamefont {S.}~\bibnamefont
  {Heusler}}, \bibinfo {author} {\bibfnamefont {S.}~\bibnamefont {M\"uller}},
  \bibinfo {author} {\bibfnamefont {A.}~\bibnamefont {Altland}}, \bibinfo
  {author} {\bibfnamefont {P.}~\bibnamefont {Braun}}, \ and\ \bibinfo {author}
  {\bibfnamefont {F.}~\bibnamefont {Haake}},\ }\href {\doibase
  10.1103/PhysRevLett.98.044103} {\bibfield  {journal} {\bibinfo  {journal}
  {Phys. Rev. Lett.}\ }\textbf {\bibinfo {volume} {98}},\ \bibinfo {pages}
  {044103} (\bibinfo {year} {2007})}\BibitemShut {NoStop}%
\bibitem [{\citenamefont {Pechukas}(1983)}]{Pechukas83}%
  \BibitemOpen
  \bibfield  {author} {\bibinfo {author} {\bibfnamefont {P.}~\bibnamefont
  {Pechukas}},\ }\href {\doibase 10.1103/PhysRevLett.51.943} {\bibfield
  {journal} {\bibinfo  {journal} {Phys. Rev. Lett.}\ }\textbf {\bibinfo
  {volume} {51}},\ \bibinfo {pages} {943} (\bibinfo {year} {1983})}\BibitemShut
  {NoStop}%
\bibitem [{\citenamefont {Yukawa}(1985)}]{Yukawa85}%
  \BibitemOpen
  \bibfield  {author} {\bibinfo {author} {\bibfnamefont {T.}~\bibnamefont
  {Yukawa}},\ }\href {\doibase 10.1103/PhysRevLett.54.1883} {\bibfield
  {journal} {\bibinfo  {journal} {Phys. Rev. Lett.}\ }\textbf {\bibinfo
  {volume} {54}},\ \bibinfo {pages} {1883} (\bibinfo {year}
  {1985})}\BibitemShut {NoStop}%
\bibitem [{\citenamefont {Haake}\ and\ \citenamefont {Ku\'s}(1988)}]{Haake88}%
  \BibitemOpen
  \bibfield  {author} {\bibinfo {author} {\bibfnamefont {F.}~\bibnamefont
  {Haake}}\ and\ \bibinfo {author} {\bibfnamefont {M.}~\bibnamefont {Ku\'s}},\
  }\href {\doibase 10.1209/0295-5075/6/7/002} {\bibfield  {journal} {\bibinfo
  {journal} {Europhysics Letters ({EPL})}\ }\textbf {\bibinfo {volume} {6}},\
  \bibinfo {pages} {579} (\bibinfo {year} {1988})}\BibitemShut {NoStop}%
\bibitem [{\citenamefont {Lenz}\ and\ \citenamefont {Haake}(1991)}]{Lenz91}%
  \BibitemOpen
  \bibfield  {author} {\bibinfo {author} {\bibfnamefont {G.}~\bibnamefont
  {Lenz}}\ and\ \bibinfo {author} {\bibfnamefont {F.}~\bibnamefont {Haake}},\
  }\href {\doibase 10.1103/PhysRevLett.67.1} {\bibfield  {journal} {\bibinfo
  {journal} {Phys. Rev. Lett.}\ }\textbf {\bibinfo {volume} {67}},\ \bibinfo
  {pages} {1} (\bibinfo {year} {1991})}\BibitemShut {NoStop}%
\bibitem [{\citenamefont {Rosenzweig}\ and\ \citenamefont
  {Porter}(1960)}]{Rosenzweig60}%
  \BibitemOpen
  \bibfield  {author} {\bibinfo {author} {\bibfnamefont {N.}~\bibnamefont
  {Rosenzweig}}\ and\ \bibinfo {author} {\bibfnamefont {C.~E.}\ \bibnamefont
  {Porter}},\ }\href {\doibase 10.1103/PhysRev.120.1698} {\bibfield  {journal}
  {\bibinfo  {journal} {Phys. Rev.}\ }\textbf {\bibinfo {volume} {120}},\
  \bibinfo {pages} {1698} (\bibinfo {year} {1960})}\BibitemShut {NoStop}%
\bibitem [{\citenamefont {Mehta}(1990)}]{Mehtabook}%
  \BibitemOpen
  \bibfield  {author} {\bibinfo {author} {\bibfnamefont {M.~L.}\ \bibnamefont
  {Mehta}},\ }\href@noop {} {\emph {\bibinfo {title} {Random Matrices}}}\
  (\bibinfo  {publisher} {Elsevier, Amsterdam},\ \bibinfo {year}
  {1990})\BibitemShut {NoStop}%
\bibitem [{\citenamefont {St\"ockmann}(1999)}]{Stockmann}%
  \BibitemOpen
  \bibfield  {author} {\bibinfo {author} {\bibfnamefont {H.}~\bibnamefont
  {St\"ockmann}},\ }\href@noop {} {\emph {\bibinfo {title} {Quantum Chaos: An
  Introduction}}}\ (\bibinfo  {publisher} {Cambridge University Press,
  Cambridge},\ \bibinfo {year} {1999})\BibitemShut {NoStop}%
\bibitem [{\citenamefont {Bohigas}\ \emph {et~al.}(1984)\citenamefont
  {Bohigas}, \citenamefont {Giannoni},\ and\ \citenamefont
  {Schmit}}]{Bohigas84}%
  \BibitemOpen
  \bibfield  {author} {\bibinfo {author} {\bibfnamefont {O.}~\bibnamefont
  {Bohigas}}, \bibinfo {author} {\bibfnamefont {M.~J.}\ \bibnamefont
  {Giannoni}}, \ and\ \bibinfo {author} {\bibfnamefont {C.}~\bibnamefont
  {Schmit}},\ }\href {\doibase 10.1103/PhysRevLett.52.1} {\bibfield  {journal}
  {\bibinfo  {journal} {Phys. Rev. Lett.}\ }\textbf {\bibinfo {volume} {52}},\
  \bibinfo {pages} {1} (\bibinfo {year} {1984})}\BibitemShut {NoStop}%
\bibitem [{\citenamefont {Nandkishore}\ and\ \citenamefont
  {Huse}(2015)}]{Nandkishore15}%
  \BibitemOpen
  \bibfield  {author} {\bibinfo {author} {\bibfnamefont {R.}~\bibnamefont
  {Nandkishore}}\ and\ \bibinfo {author} {\bibfnamefont {D.~A.}\ \bibnamefont
  {Huse}},\ }\href {\doibase 10.1146/annurev-conmatphys-031214-014726}
  {\bibfield  {journal} {\bibinfo  {journal} {Annual Review of Condensed Matter
  Physics}\ }\textbf {\bibinfo {volume} {6}},\ \bibinfo {pages} {15} (\bibinfo
  {year} {2015})},\ \Eprint
  {http://arxiv.org/abs/https://doi.org/10.1146/annurev-conmatphys-031214-014726}
  {https://doi.org/10.1146/annurev-conmatphys-031214-014726} \BibitemShut
  {NoStop}%
\bibitem [{\citenamefont {Alet}\ and\ \citenamefont
  {Laflorencie}(2018)}]{Alet18}%
  \BibitemOpen
  \bibfield  {author} {\bibinfo {author} {\bibfnamefont {F.}~\bibnamefont
  {Alet}}\ and\ \bibinfo {author} {\bibfnamefont {N.}~\bibnamefont
  {Laflorencie}},\ }\href {\doibase https://doi.org/10.1016/j.crhy.2018.03.003}
  {\bibfield  {journal} {\bibinfo  {journal} {Comptes Rendus Physique}\
  }\textbf {\bibinfo {volume} {19}},\ \bibinfo {pages} {498 } (\bibinfo {year}
  {2018})},\ \bibinfo {note} {quantum simulation / Simulation
  quantique}\BibitemShut {NoStop}%
\bibitem [{\citenamefont {Abanin}\ \emph {et~al.}(2019)\citenamefont {Abanin},
  \citenamefont {Altman}, \citenamefont {Bloch},\ and\ \citenamefont
  {Serbyn}}]{Abanin19}%
  \BibitemOpen
  \bibfield  {author} {\bibinfo {author} {\bibfnamefont {D.~A.}\ \bibnamefont
  {Abanin}}, \bibinfo {author} {\bibfnamefont {E.}~\bibnamefont {Altman}},
  \bibinfo {author} {\bibfnamefont {I.}~\bibnamefont {Bloch}}, \ and\ \bibinfo
  {author} {\bibfnamefont {M.}~\bibnamefont {Serbyn}},\ }\href {\doibase
  10.1103/RevModPhys.91.021001} {\bibfield  {journal} {\bibinfo  {journal}
  {Rev. Mod. Phys.}\ }\textbf {\bibinfo {volume} {91}},\ \bibinfo {pages}
  {021001} (\bibinfo {year} {2019})}\BibitemShut {NoStop}%
\bibitem [{\citenamefont {Brody}(1973)}]{Brody73}%
  \BibitemOpen
  \bibfield  {author} {\bibinfo {author} {\bibfnamefont {T.~A.}\ \bibnamefont
  {Brody}},\ }\href {\doibase 10.1007/BF02727859} {\bibfield  {journal}
  {\bibinfo  {journal} {Lettere al Nuovo Cimento (1971-1985)}\ }\textbf
  {\bibinfo {volume} {7}},\ \bibinfo {pages} {482} (\bibinfo {year}
  {1973})}\BibitemShut {NoStop}%
\bibitem [{\citenamefont {Berry}\ and\ \citenamefont
  {Robnik}(1984)}]{Berry84b}%
  \BibitemOpen
  \bibfield  {author} {\bibinfo {author} {\bibfnamefont {M.~V.}\ \bibnamefont
  {Berry}}\ and\ \bibinfo {author} {\bibfnamefont {M.}~\bibnamefont {Robnik}},\
  }\href {\doibase 10.1088/0305-4470/17/12/013} {\bibfield  {journal} {\bibinfo
   {journal} {Journal of Physics A: Mathematical and General}\ }\textbf
  {\bibinfo {volume} {17}},\ \bibinfo {pages} {2413} (\bibinfo {year}
  {1984})}\BibitemShut {NoStop}%
\bibitem [{\citenamefont {Prosen}(1998)}]{Prosen98}%
  \BibitemOpen
  \bibfield  {author} {\bibinfo {author} {\bibfnamefont {T.}~\bibnamefont
  {Prosen}},\ }\href {\doibase 10.1088/0305-4470/31/34/005} {\bibfield
  {journal} {\bibinfo  {journal} {Journal of Physics A: Mathematical and
  General}\ }\textbf {\bibinfo {volume} {31}},\ \bibinfo {pages} {7023}
  (\bibinfo {year} {1998})}\BibitemShut {NoStop}%
\bibitem [{\citenamefont {Prosen}\ and\ \citenamefont
  {Robnik}(1999)}]{Prosen99}%
  \BibitemOpen
  \bibfield  {author} {\bibinfo {author} {\bibfnamefont {T.}~\bibnamefont
  {Prosen}}\ and\ \bibinfo {author} {\bibfnamefont {M.}~\bibnamefont
  {Robnik}},\ }\href {\doibase 10.1088/0305-4470/32/10/006} {\bibfield
  {journal} {\bibinfo  {journal} {J. Phys. A: Math. Gen.}\ }\textbf {\bibinfo
  {volume} {32}},\ \bibinfo {pages} {1863} (\bibinfo {year}
  {1999})}\BibitemShut {NoStop}%
\bibitem [{\citenamefont {Khaymovich}\ \emph {et~al.}(2020)\citenamefont
  {Khaymovich}, \citenamefont {Kravtsov}, \citenamefont {Altshuler},\ and\
  \citenamefont {Ioffe}}]{Khaymovich20}%
  \BibitemOpen
  \bibfield  {author} {\bibinfo {author} {\bibfnamefont {I.~M.}\ \bibnamefont
  {Khaymovich}}, \bibinfo {author} {\bibfnamefont {V.~E.}\ \bibnamefont
  {Kravtsov}}, \bibinfo {author} {\bibfnamefont {B.~L.}\ \bibnamefont
  {Altshuler}}, \ and\ \bibinfo {author} {\bibfnamefont {L.~B.}\ \bibnamefont
  {Ioffe}},\ }\href {\doibase 10.1103/PhysRevResearch.2.043346} {\bibfield
  {journal} {\bibinfo  {journal} {Phys. Rev. Research}\ }\textbf {\bibinfo
  {volume} {2}},\ \bibinfo {pages} {043346} (\bibinfo {year}
  {2020})}\BibitemShut {NoStop}%
\bibitem [{\citenamefont {Kravtsov}\ \emph {et~al.}(2020)\citenamefont
  {Kravtsov}, \citenamefont {Khaymovich}, \citenamefont {Altshuler},\ and\
  \citenamefont {Ioffe}}]{Kravtsov20}%
  \BibitemOpen
  \bibfield  {author} {\bibinfo {author} {\bibfnamefont {V.}~\bibnamefont
  {Kravtsov}}, \bibinfo {author} {\bibfnamefont {I.}~\bibnamefont
  {Khaymovich}}, \bibinfo {author} {\bibfnamefont {B.}~\bibnamefont
  {Altshuler}}, \ and\ \bibinfo {author} {\bibfnamefont {L.}~\bibnamefont
  {Ioffe}},\ }\href@noop {} {\bibfield  {journal} {\bibinfo  {journal} {arXiv
  preprint arXiv:2002.02979}\ } (\bibinfo {year} {2020})}\BibitemShut {NoStop}%
\bibitem [{\citenamefont {Khaymovich}\ and\ \citenamefont
  {Kravtsov}(2021)}]{Khaymovich21}%
  \BibitemOpen
  \bibfield  {author} {\bibinfo {author} {\bibfnamefont {I.}~\bibnamefont
  {Khaymovich}}\ and\ \bibinfo {author} {\bibfnamefont {V.}~\bibnamefont
  {Kravtsov}},\ }\href@noop {} {\bibfield  {journal} {\bibinfo  {journal}
  {arXiv preprint arXiv:2106.01965}\ } (\bibinfo {year} {2021})}\BibitemShut
  {NoStop}%
\bibitem [{\citenamefont {Seligman}\ \emph {et~al.}(1984)\citenamefont
  {Seligman}, \citenamefont {Verbaarschot},\ and\ \citenamefont
  {Zirnbauer}}]{Seligman84}%
  \BibitemOpen
  \bibfield  {author} {\bibinfo {author} {\bibfnamefont {T.~H.}\ \bibnamefont
  {Seligman}}, \bibinfo {author} {\bibfnamefont {J.~J.~M.}\ \bibnamefont
  {Verbaarschot}}, \ and\ \bibinfo {author} {\bibfnamefont {M.~R.}\
  \bibnamefont {Zirnbauer}},\ }\href {\doibase 10.1103/PhysRevLett.53.215}
  {\bibfield  {journal} {\bibinfo  {journal} {Phys. Rev. Lett.}\ }\textbf
  {\bibinfo {volume} {53}},\ \bibinfo {pages} {215} (\bibinfo {year}
  {1984})}\BibitemShut {NoStop}%
\bibitem [{\citenamefont {Guhr}(1996)}]{Guhr96}%
  \BibitemOpen
  \bibfield  {author} {\bibinfo {author} {\bibfnamefont {T.}~\bibnamefont
  {Guhr}},\ }\href {\doibase https://doi.org/10.1006/aphy.1996.0091} {\bibfield
   {journal} {\bibinfo  {journal} {Annals of Physics}\ }\textbf {\bibinfo
  {volume} {250}},\ \bibinfo {pages} {145} (\bibinfo {year}
  {1996})}\BibitemShut {NoStop}%
\bibitem [{\citenamefont {Casati}\ \emph {et~al.}(1990)\citenamefont {Casati},
  \citenamefont {Molinari},\ and\ \citenamefont {Izrailev}}]{Casati90}%
  \BibitemOpen
  \bibfield  {author} {\bibinfo {author} {\bibfnamefont {G.}~\bibnamefont
  {Casati}}, \bibinfo {author} {\bibfnamefont {L.}~\bibnamefont {Molinari}}, \
  and\ \bibinfo {author} {\bibfnamefont {F.}~\bibnamefont {Izrailev}},\ }\href
  {\doibase 10.1103/PhysRevLett.64.1851} {\bibfield  {journal} {\bibinfo
  {journal} {Phys. Rev. Lett.}\ }\textbf {\bibinfo {volume} {64}},\ \bibinfo
  {pages} {1851} (\bibinfo {year} {1990})}\BibitemShut {NoStop}%
\bibitem [{\citenamefont {Casati}\ \emph {et~al.}(1991)\citenamefont {Casati},
  \citenamefont {Izrailev},\ and\ \citenamefont {Molinari}}]{Casati91}%
  \BibitemOpen
  \bibfield  {author} {\bibinfo {author} {\bibfnamefont {G.}~\bibnamefont
  {Casati}}, \bibinfo {author} {\bibfnamefont {F.}~\bibnamefont {Izrailev}}, \
  and\ \bibinfo {author} {\bibfnamefont {L.}~\bibnamefont {Molinari}},\ }\href
  {\doibase 10.1088/0305-4470/24/20/011} {\bibfield  {journal} {\bibinfo
  {journal} {Journal of Physics A: Mathematical and General}\ }\textbf
  {\bibinfo {volume} {24}},\ \bibinfo {pages} {4755} (\bibinfo {year}
  {1991})}\BibitemShut {NoStop}%
\bibitem [{\citenamefont {Fyodorov}\ and\ \citenamefont
  {Mirlin}(1991)}]{fyodorov91}%
  \BibitemOpen
  \bibfield  {author} {\bibinfo {author} {\bibfnamefont {Y.~V.}\ \bibnamefont
  {Fyodorov}}\ and\ \bibinfo {author} {\bibfnamefont {A.~D.}\ \bibnamefont
  {Mirlin}},\ }\href {\doibase 10.1103/PhysRevLett.67.2405} {\bibfield
  {journal} {\bibinfo  {journal} {Phys. Rev. Lett.}\ }\textbf {\bibinfo
  {volume} {67}},\ \bibinfo {pages} {2405} (\bibinfo {year}
  {1991})}\BibitemShut {NoStop}%
\bibitem [{\citenamefont {{Bogomolny, E.}}\ \emph {et~al.}(2001)\citenamefont
  {{Bogomolny, E.}}, \citenamefont {{Gerland, U.}},\ and\ \citenamefont
  {{Schmit, C.}}}]{Bogomolny01}%
  \BibitemOpen
  \bibfield  {author} {\bibinfo {author} {\bibnamefont {{Bogomolny, E.}}},
  \bibinfo {author} {\bibnamefont {{Gerland, U.}}}, \ and\ \bibinfo {author}
  {\bibnamefont {{Schmit, C.}}},\ }\href {\doibase 10.1007/s100510170357}
  {\bibfield  {journal} {\bibinfo  {journal} {Eur. Phys. J. B}\ }\textbf
  {\bibinfo {volume} {19}},\ \bibinfo {pages} {121} (\bibinfo {year}
  {2001})}\BibitemShut {NoStop}%
\bibitem [{\citenamefont {Bogomolny}\ and\ \citenamefont
  {Giraud}(2011)}]{Bogomolny11}%
  \BibitemOpen
  \bibfield  {author} {\bibinfo {author} {\bibfnamefont {E.}~\bibnamefont
  {Bogomolny}}\ and\ \bibinfo {author} {\bibfnamefont {O.}~\bibnamefont
  {Giraud}},\ }\href {\doibase 10.1103/PhysRevLett.106.044101} {\bibfield
  {journal} {\bibinfo  {journal} {Phys. Rev. Lett.}\ }\textbf {\bibinfo
  {volume} {106}},\ \bibinfo {pages} {044101} (\bibinfo {year}
  {2011})}\BibitemShut {NoStop}%
\bibitem [{\citenamefont {Bertrand}\ and\ \citenamefont
  {Garc\'{\i}a-Garc\'{\i}a}(2016)}]{Garcia16}%
  \BibitemOpen
  \bibfield  {author} {\bibinfo {author} {\bibfnamefont {C.~L.}\ \bibnamefont
  {Bertrand}}\ and\ \bibinfo {author} {\bibfnamefont {A.~M.}\ \bibnamefont
  {Garc\'{\i}a-Garc\'{\i}a}},\ }\href {\doibase 10.1103/PhysRevB.94.144201}
  {\bibfield  {journal} {\bibinfo  {journal} {Phys. Rev. B}\ }\textbf {\bibinfo
  {volume} {94}},\ \bibinfo {pages} {144201} (\bibinfo {year}
  {2016})}\BibitemShut {NoStop}%
\bibitem [{\citenamefont {G\'omez}\ \emph {et~al.}(2002)\citenamefont
  {G\'omez}, \citenamefont {Molina}, \citenamefont {Rela\~no},\ and\
  \citenamefont {Retamosa}}]{Gomez02}%
  \BibitemOpen
  \bibfield  {author} {\bibinfo {author} {\bibfnamefont {J.~M.~G.}\
  \bibnamefont {G\'omez}}, \bibinfo {author} {\bibfnamefont {R.~A.}\
  \bibnamefont {Molina}}, \bibinfo {author} {\bibfnamefont {A.}~\bibnamefont
  {Rela\~no}}, \ and\ \bibinfo {author} {\bibfnamefont {J.}~\bibnamefont
  {Retamosa}},\ }\href {\doibase 10.1103/PhysRevE.66.036209} {\bibfield
  {journal} {\bibinfo  {journal} {Phys. Rev. E}\ }\textbf {\bibinfo {volume}
  {66}},\ \bibinfo {pages} {036209} (\bibinfo {year} {2002})}\BibitemShut
  {NoStop}%
\bibitem [{\citenamefont {Morales}\ \emph {et~al.}(2011)\citenamefont
  {Morales}, \citenamefont {Landa}, \citenamefont {Str\'ansk\'y},\ and\
  \citenamefont {Frank}}]{Irving}%
  \BibitemOpen
  \bibfield  {author} {\bibinfo {author} {\bibfnamefont {I.~O.}\ \bibnamefont
  {Morales}}, \bibinfo {author} {\bibfnamefont {E.}~\bibnamefont {Landa}},
  \bibinfo {author} {\bibfnamefont {P.}~\bibnamefont {Str\'ansk\'y}}, \ and\
  \bibinfo {author} {\bibfnamefont {A.}~\bibnamefont {Frank}},\ }\href
  {\doibase 10.1103/PhysRevE.84.016203} {\bibfield  {journal} {\bibinfo
  {journal} {Phys. Rev. E}\ }\textbf {\bibinfo {volume} {84}},\ \bibinfo
  {pages} {016203} (\bibinfo {year} {2011})}\BibitemShut {NoStop}%
\bibitem [{\citenamefont {Torres-Vargas}\ \emph {et~al.}(2017)\citenamefont
  {Torres-Vargas}, \citenamefont {Fossion}, \citenamefont {Tapia-Ignacio},\
  and\ \citenamefont {L\'opez-Vieyra}}]{Vargas}%
  \BibitemOpen
  \bibfield  {author} {\bibinfo {author} {\bibfnamefont {G.}~\bibnamefont
  {Torres-Vargas}}, \bibinfo {author} {\bibfnamefont {R.}~\bibnamefont
  {Fossion}}, \bibinfo {author} {\bibfnamefont {C.}~\bibnamefont
  {Tapia-Ignacio}}, \ and\ \bibinfo {author} {\bibfnamefont {J.~C.}\
  \bibnamefont {L\'opez-Vieyra}},\ }\href {\doibase 10.1103/PhysRevE.96.012110}
  {\bibfield  {journal} {\bibinfo  {journal} {Phys. Rev. E}\ }\textbf {\bibinfo
  {volume} {96}},\ \bibinfo {pages} {012110} (\bibinfo {year}
  {2017})}\BibitemShut {NoStop}%
\bibitem [{\citenamefont {Oganesyan}\ and\ \citenamefont
  {Huse}(2007)}]{Oganesyan07}%
  \BibitemOpen
  \bibfield  {author} {\bibinfo {author} {\bibfnamefont {V.}~\bibnamefont
  {Oganesyan}}\ and\ \bibinfo {author} {\bibfnamefont {D.~A.}\ \bibnamefont
  {Huse}},\ }\href {\doibase 10.1103/PhysRevB.75.155111} {\bibfield  {journal}
  {\bibinfo  {journal} {Phys. Rev. B}\ }\textbf {\bibinfo {volume} {75}},\
  \bibinfo {pages} {155111} (\bibinfo {year} {2007})}\BibitemShut {NoStop}%
\bibitem [{\citenamefont {Santos}\ \emph {et~al.}(2004)\citenamefont {Santos},
  \citenamefont {Rigolin},\ and\ \citenamefont {Escobar}}]{Santos04}%
  \BibitemOpen
  \bibfield  {author} {\bibinfo {author} {\bibfnamefont {L.~F.}\ \bibnamefont
  {Santos}}, \bibinfo {author} {\bibfnamefont {G.}~\bibnamefont {Rigolin}}, \
  and\ \bibinfo {author} {\bibfnamefont {C.~O.}\ \bibnamefont {Escobar}},\
  }\href {\doibase 10.1103/PhysRevA.69.042304} {\bibfield  {journal} {\bibinfo
  {journal} {Phys. Rev. A}\ }\textbf {\bibinfo {volume} {69}},\ \bibinfo
  {pages} {042304} (\bibinfo {year} {2004})}\BibitemShut {NoStop}%
\bibitem [{\citenamefont {Atas}\ \emph {et~al.}(2013)\citenamefont {Atas},
  \citenamefont {Bogomolny}, \citenamefont {Giraud},\ and\ \citenamefont
  {Roux}}]{Atas13}%
  \BibitemOpen
  \bibfield  {author} {\bibinfo {author} {\bibfnamefont {Y.~Y.}\ \bibnamefont
  {Atas}}, \bibinfo {author} {\bibfnamefont {E.}~\bibnamefont {Bogomolny}},
  \bibinfo {author} {\bibfnamefont {O.}~\bibnamefont {Giraud}}, \ and\ \bibinfo
  {author} {\bibfnamefont {G.}~\bibnamefont {Roux}},\ }\href {\doibase
  10.1103/PhysRevLett.110.084101} {\bibfield  {journal} {\bibinfo  {journal}
  {Phys. Rev. Lett.}\ }\textbf {\bibinfo {volume} {110}},\ \bibinfo {pages}
  {084101} (\bibinfo {year} {2013})}\BibitemShut {NoStop}%
\bibitem [{\citenamefont {Santos}\ and\ \citenamefont
  {Rigol}(2010)}]{Santos10}%
  \BibitemOpen
  \bibfield  {author} {\bibinfo {author} {\bibfnamefont {L.~F.}\ \bibnamefont
  {Santos}}\ and\ \bibinfo {author} {\bibfnamefont {M.}~\bibnamefont {Rigol}},\
  }\href {\doibase 10.1103/PhysRevE.81.036206} {\bibfield  {journal} {\bibinfo
  {journal} {Phys. Rev. E}\ }\textbf {\bibinfo {volume} {81}},\ \bibinfo
  {pages} {036206} (\bibinfo {year} {2010})}\BibitemShut {NoStop}%
\bibitem [{\citenamefont {Pal}\ and\ \citenamefont {Huse}(2010)}]{Pal10}%
  \BibitemOpen
  \bibfield  {author} {\bibinfo {author} {\bibfnamefont {A.}~\bibnamefont
  {Pal}}\ and\ \bibinfo {author} {\bibfnamefont {D.~A.}\ \bibnamefont {Huse}},\
  }\href {\doibase 10.1103/PhysRevB.82.174411} {\bibfield  {journal} {\bibinfo
  {journal} {Phys. Rev. B}\ }\textbf {\bibinfo {volume} {82}},\ \bibinfo
  {pages} {174411} (\bibinfo {year} {2010})}\BibitemShut {NoStop}%
\bibitem [{\citenamefont {Mondaini}\ and\ \citenamefont
  {Rigol}(2015)}]{Mondaini15}%
  \BibitemOpen
  \bibfield  {author} {\bibinfo {author} {\bibfnamefont {R.}~\bibnamefont
  {Mondaini}}\ and\ \bibinfo {author} {\bibfnamefont {M.}~\bibnamefont
  {Rigol}},\ }\href {\doibase 10.1103/PhysRevA.92.041601} {\bibfield  {journal}
  {\bibinfo  {journal} {Phys. Rev. A}\ }\textbf {\bibinfo {volume} {92}},\
  \bibinfo {pages} {041601} (\bibinfo {year} {2015})}\BibitemShut {NoStop}%
\bibitem [{\citenamefont {Luitz}\ \emph {et~al.}(2015)\citenamefont {Luitz},
  \citenamefont {Laflorencie},\ and\ \citenamefont {Alet}}]{Luitz15}%
  \BibitemOpen
  \bibfield  {author} {\bibinfo {author} {\bibfnamefont {D.~J.}\ \bibnamefont
  {Luitz}}, \bibinfo {author} {\bibfnamefont {N.}~\bibnamefont {Laflorencie}},
  \ and\ \bibinfo {author} {\bibfnamefont {F.}~\bibnamefont {Alet}},\ }\href
  {\doibase 10.1103/PhysRevB.91.081103} {\bibfield  {journal} {\bibinfo
  {journal} {Phys. Rev. B}\ }\textbf {\bibinfo {volume} {91}},\ \bibinfo
  {pages} {081103} (\bibinfo {year} {2015})}\BibitemShut {NoStop}%
\bibitem [{\citenamefont {Serbyn}\ and\ \citenamefont
  {Moore}(2016)}]{Serbyn16}%
  \BibitemOpen
  \bibfield  {author} {\bibinfo {author} {\bibfnamefont {M.}~\bibnamefont
  {Serbyn}}\ and\ \bibinfo {author} {\bibfnamefont {J.~E.}\ \bibnamefont
  {Moore}},\ }\href {\doibase 10.1103/PhysRevB.93.041424} {\bibfield  {journal}
  {\bibinfo  {journal} {Phys. Rev. B}\ }\textbf {\bibinfo {volume} {93}},\
  \bibinfo {pages} {041424} (\bibinfo {year} {2016})}\BibitemShut {NoStop}%
\bibitem [{\citenamefont {Kravtsov}\ and\ \citenamefont
  {Lerner}(1995)}]{Kravtsov95}%
  \BibitemOpen
  \bibfield  {author} {\bibinfo {author} {\bibfnamefont {V.~E.}\ \bibnamefont
  {Kravtsov}}\ and\ \bibinfo {author} {\bibfnamefont {I.~V.}\ \bibnamefont
  {Lerner}},\ }\href {http://stacks.iop.org/0305-4470/28/i=13/a=008} {\bibfield
   {journal} {\bibinfo  {journal} {J. Phys. A: Mat. Gen.}\ }\textbf {\bibinfo
  {volume} {28}},\ \bibinfo {pages} {3623} (\bibinfo {year}
  {1995})}\BibitemShut {NoStop}%
\bibitem [{\citenamefont {Bogomolny}\ \emph {et~al.}(1999)\citenamefont
  {Bogomolny}, \citenamefont {Gerland},\ and\ \citenamefont
  {Schmit}}]{Bogomolny99}%
  \BibitemOpen
  \bibfield  {author} {\bibinfo {author} {\bibfnamefont {E.~B.}\ \bibnamefont
  {Bogomolny}}, \bibinfo {author} {\bibfnamefont {U.}~\bibnamefont {Gerland}},
  \ and\ \bibinfo {author} {\bibfnamefont {C.}~\bibnamefont {Schmit}},\ }\href
  {\doibase 10.1103/PhysRevE.59.R1315} {\bibfield  {journal} {\bibinfo
  {journal} {Phys. Rev. E}\ }\textbf {\bibinfo {volume} {59}},\ \bibinfo
  {pages} {R1315} (\bibinfo {year} {1999})}\BibitemShut {NoStop}%
\bibitem [{\citenamefont {Buijsman}\ \emph {et~al.}(2019)\citenamefont
  {Buijsman}, \citenamefont {Cheianov},\ and\ \citenamefont
  {Gritsev}}]{Buijsman18}%
  \BibitemOpen
  \bibfield  {author} {\bibinfo {author} {\bibfnamefont {W.}~\bibnamefont
  {Buijsman}}, \bibinfo {author} {\bibfnamefont {V.}~\bibnamefont {Cheianov}},
  \ and\ \bibinfo {author} {\bibfnamefont {V.}~\bibnamefont {Gritsev}},\ }\href
  {\doibase 10.1103/PhysRevLett.122.180601} {\bibfield  {journal} {\bibinfo
  {journal} {Phys. Rev. Lett.}\ }\textbf {\bibinfo {volume} {122}},\ \bibinfo
  {pages} {180601} (\bibinfo {year} {2019})}\BibitemShut {NoStop}%
\bibitem [{\citenamefont {Sierant}\ and\ \citenamefont
  {Zakrzewski}(2019)}]{Sierant19b}%
  \BibitemOpen
  \bibfield  {author} {\bibinfo {author} {\bibfnamefont {P.}~\bibnamefont
  {Sierant}}\ and\ \bibinfo {author} {\bibfnamefont {J.}~\bibnamefont
  {Zakrzewski}},\ }\href {\doibase 10.1103/PhysRevB.99.104205} {\bibfield
  {journal} {\bibinfo  {journal} {Phys. Rev. B}\ }\textbf {\bibinfo {volume}
  {99}},\ \bibinfo {pages} {104205} (\bibinfo {year} {2019})}\BibitemShut
  {NoStop}%
\bibitem [{\citenamefont {Sierant}\ and\ \citenamefont
  {Zakrzewski}(2020)}]{Sierant20}%
  \BibitemOpen
  \bibfield  {author} {\bibinfo {author} {\bibfnamefont {P.}~\bibnamefont
  {Sierant}}\ and\ \bibinfo {author} {\bibfnamefont {J.}~\bibnamefont
  {Zakrzewski}},\ }\href {\doibase 10.1103/PhysRevB.101.104201} {\bibfield
  {journal} {\bibinfo  {journal} {Phys. Rev. B}\ }\textbf {\bibinfo {volume}
  {101}},\ \bibinfo {pages} {104201} (\bibinfo {year} {2020})}\BibitemShut
  {NoStop}%
\bibitem [{\citenamefont {Chavda}\ \emph {et~al.}(2014)\citenamefont {Chavda},
  \citenamefont {Deota},\ and\ \citenamefont {Kota}}]{Chavda14}%
  \BibitemOpen
  \bibfield  {author} {\bibinfo {author} {\bibfnamefont {N.}~\bibnamefont
  {Chavda}}, \bibinfo {author} {\bibfnamefont {H.~N.}\ \bibnamefont {Deota}}, \
  and\ \bibinfo {author} {\bibfnamefont {V.~K.~B.}\ \bibnamefont {Kota}},\
  }\href {\doibase 10.1016/j.physleta.2014.08.021} {\bibfield  {journal}
  {\bibinfo  {journal} {Phys. Lett. A}\ }\textbf {\bibinfo {volume} {378}},\
  \bibinfo {pages} {3012} (\bibinfo {year} {2014})}\BibitemShut {NoStop}%
\bibitem [{\citenamefont {Tekur}\ \emph {et~al.}(2018)\citenamefont {Tekur},
  \citenamefont {Bhosale},\ and\ \citenamefont {Santhanam}}]{Tekur18}%
  \BibitemOpen
  \bibfield  {author} {\bibinfo {author} {\bibfnamefont {S.~H.}\ \bibnamefont
  {Tekur}}, \bibinfo {author} {\bibfnamefont {U.~T.}\ \bibnamefont {Bhosale}},
  \ and\ \bibinfo {author} {\bibfnamefont {M.~S.}\ \bibnamefont {Santhanam}},\
  }\href {\doibase 10.1103/PhysRevB.98.104305} {\bibfield  {journal} {\bibinfo
  {journal} {Phys. Rev. B}\ }\textbf {\bibinfo {volume} {98}},\ \bibinfo
  {pages} {104305} (\bibinfo {year} {2018})}\BibitemShut {NoStop}%
\bibitem [{\citenamefont {Zakrzewski}\ and\ \citenamefont
  {Delande}(1993)}]{Zakrzewski93}%
  \BibitemOpen
  \bibfield  {author} {\bibinfo {author} {\bibfnamefont {J.}~\bibnamefont
  {Zakrzewski}}\ and\ \bibinfo {author} {\bibfnamefont {D.}~\bibnamefont
  {Delande}},\ }\href {\doibase 10.1103/PhysRevE.47.1650} {\bibfield  {journal}
  {\bibinfo  {journal} {Phys. Rev. E}\ }\textbf {\bibinfo {volume} {47}},\
  \bibinfo {pages} {1650} (\bibinfo {year} {1993})}\BibitemShut {NoStop}%
\bibitem [{\citenamefont {Zakrzewski}\ \emph {et~al.}(1993)\citenamefont
  {Zakrzewski}, \citenamefont {Delande},\ and\ \citenamefont
  {Ku\ifmmode~\acute{s}\else \'{s}\fi{}}}]{Zakrzewski93b}%
  \BibitemOpen
  \bibfield  {author} {\bibinfo {author} {\bibfnamefont {J.}~\bibnamefont
  {Zakrzewski}}, \bibinfo {author} {\bibfnamefont {D.}~\bibnamefont {Delande}},
  \ and\ \bibinfo {author} {\bibfnamefont {M.}~\bibnamefont
  {Ku\ifmmode~\acute{s}\else \'{s}\fi{}}},\ }\href {\doibase
  10.1103/PhysRevE.47.1665} {\bibfield  {journal} {\bibinfo  {journal} {Phys.
  Rev. E}\ }\textbf {\bibinfo {volume} {47}},\ \bibinfo {pages} {1665}
  (\bibinfo {year} {1993})}\BibitemShut {NoStop}%
\bibitem [{\citenamefont {Vidmar}\ and\ \citenamefont
  {Rigol}(2016)}]{Vidmar16}%
  \BibitemOpen
  \bibfield  {author} {\bibinfo {author} {\bibfnamefont {L.}~\bibnamefont
  {Vidmar}}\ and\ \bibinfo {author} {\bibfnamefont {M.}~\bibnamefont {Rigol}},\
  }\href {\doibase 10.1088/1742-5468/2016/06/064007} {\bibfield  {journal}
  {\bibinfo  {journal} {Journal of Statistical Mechanics: Theory and
  Experiment}\ }\textbf {\bibinfo {volume} {2016}},\ \bibinfo {pages} {064007}
  (\bibinfo {year} {2016})}\BibitemShut {NoStop}%
\bibitem [{\citenamefont {Metropolis}\ \emph {et~al.}(1953)\citenamefont
  {Metropolis}, \citenamefont {Rosenbluth}, \citenamefont {Rosenbluth},
  \citenamefont {Teller},\ and\ \citenamefont {Teller}}]{Metropolis53}%
  \BibitemOpen
  \bibfield  {author} {\bibinfo {author} {\bibfnamefont {N.}~\bibnamefont
  {Metropolis}}, \bibinfo {author} {\bibfnamefont {A.~W.}\ \bibnamefont
  {Rosenbluth}}, \bibinfo {author} {\bibfnamefont {M.~N.}\ \bibnamefont
  {Rosenbluth}}, \bibinfo {author} {\bibfnamefont {A.~H.}\ \bibnamefont
  {Teller}}, \ and\ \bibinfo {author} {\bibfnamefont {E.}~\bibnamefont
  {Teller}},\ }\href {\doibase 10.1063/1.1699114} {\bibfield  {journal}
  {\bibinfo  {journal} {The Journal of Chemical Physics}\ }\textbf {\bibinfo
  {volume} {21}},\ \bibinfo {pages} {1087} (\bibinfo {year}
  {1953})}\BibitemShut {NoStop}%
\bibitem [{\citenamefont {Hastings}(1970)}]{Hastings70}%
  \BibitemOpen
  \bibfield  {author} {\bibinfo {author} {\bibfnamefont {W.~K.}\ \bibnamefont
  {Hastings}},\ }\href {\doibase 10.1093/biomet/57.1.97} {\bibfield  {journal}
  {\bibinfo  {journal} {Biometrika}\ }\textbf {\bibinfo {volume} {57}},\
  \bibinfo {pages} {97} (\bibinfo {year} {1970})}\BibitemShut {NoStop}%
\bibitem [{\citenamefont {Luitz}\ \emph {et~al.}(2016)\citenamefont {Luitz},
  \citenamefont {Laflorencie},\ and\ \citenamefont {Alet}}]{Luitz16}%
  \BibitemOpen
  \bibfield  {author} {\bibinfo {author} {\bibfnamefont {D.~J.}\ \bibnamefont
  {Luitz}}, \bibinfo {author} {\bibfnamefont {N.}~\bibnamefont {Laflorencie}},
  \ and\ \bibinfo {author} {\bibfnamefont {F.}~\bibnamefont {Alet}},\ }\href
  {\doibase 10.1103/PhysRevB.93.060201} {\bibfield  {journal} {\bibinfo
  {journal} {Phys. Rev. B}\ }\textbf {\bibinfo {volume} {93}},\ \bibinfo
  {pages} {060201} (\bibinfo {year} {2016})}\BibitemShut {NoStop}%
\bibitem [{\citenamefont {Mac\'e}\ \emph {et~al.}(2019)\citenamefont {Mac\'e},
  \citenamefont {Laflorencie},\ and\ \citenamefont {Alet}}]{Mace19}%
  \BibitemOpen
  \bibfield  {author} {\bibinfo {author} {\bibfnamefont {N.}~\bibnamefont
  {Mac\'e}}, \bibinfo {author} {\bibfnamefont {N.}~\bibnamefont {Laflorencie}},
  \ and\ \bibinfo {author} {\bibfnamefont {F.}~\bibnamefont {Alet}},\ }\href
  {\doibase 10.21468/SciPostPhys.6.4.050} {\bibfield  {journal} {\bibinfo
  {journal} {SciPost Phys.}\ }\textbf {\bibinfo {volume} {6}},\ \bibinfo
  {pages} {50} (\bibinfo {year} {2019})}\BibitemShut {NoStop}%
\bibitem [{\citenamefont {Guarrera}\ \emph {et~al.}(2007)\citenamefont
  {Guarrera}, \citenamefont {Fallani}, \citenamefont {Lye}, \citenamefont
  {Fort},\ and\ \citenamefont {Inguscio}}]{Guarrera07}%
  \BibitemOpen
  \bibfield  {author} {\bibinfo {author} {\bibfnamefont {V.}~\bibnamefont
  {Guarrera}}, \bibinfo {author} {\bibfnamefont {L.}~\bibnamefont {Fallani}},
  \bibinfo {author} {\bibfnamefont {J.~E.}\ \bibnamefont {Lye}}, \bibinfo
  {author} {\bibfnamefont {C.}~\bibnamefont {Fort}}, \ and\ \bibinfo {author}
  {\bibfnamefont {M.}~\bibnamefont {Inguscio}},\ }\href {\doibase
  10.1088/1367-2630/9/4/107} {\bibfield  {journal} {\bibinfo  {journal} {New J.
  Phys.}\ }\textbf {\bibinfo {volume} {9}},\ \bibinfo {pages} {107} (\bibinfo
  {year} {2007})}\BibitemShut {NoStop}%
\bibitem [{\citenamefont {Doggen}\ and\ \citenamefont
  {Mirlin}(2019)}]{Doggen19}%
  \BibitemOpen
  \bibfield  {author} {\bibinfo {author} {\bibfnamefont {E.~V.~H.}\
  \bibnamefont {Doggen}}\ and\ \bibinfo {author} {\bibfnamefont {A.~D.}\
  \bibnamefont {Mirlin}},\ }\href {\doibase 10.1103/PhysRevB.100.104203}
  {\bibfield  {journal} {\bibinfo  {journal} {Phys. Rev. B}\ }\textbf {\bibinfo
  {volume} {100}},\ \bibinfo {pages} {104203} (\bibinfo {year}
  {2019})}\BibitemShut {NoStop}%
\bibitem [{\citenamefont {Schreiber}\ \emph {et~al.}(2015)\citenamefont
  {Schreiber}, \citenamefont {Hodgman}, \citenamefont {Bordia}, \citenamefont
  {L{\"u}schen}, \citenamefont {Fischer}, \citenamefont {Vosk}, \citenamefont
  {Altman}, \citenamefont {Schneider},\ and\ \citenamefont
  {Bloch}}]{Schreiber15}%
  \BibitemOpen
  \bibfield  {author} {\bibinfo {author} {\bibfnamefont {M.}~\bibnamefont
  {Schreiber}}, \bibinfo {author} {\bibfnamefont {S.~S.}\ \bibnamefont
  {Hodgman}}, \bibinfo {author} {\bibfnamefont {P.}~\bibnamefont {Bordia}},
  \bibinfo {author} {\bibfnamefont {H.~P.}\ \bibnamefont {L{\"u}schen}},
  \bibinfo {author} {\bibfnamefont {M.~H.}\ \bibnamefont {Fischer}}, \bibinfo
  {author} {\bibfnamefont {R.}~\bibnamefont {Vosk}}, \bibinfo {author}
  {\bibfnamefont {E.}~\bibnamefont {Altman}}, \bibinfo {author} {\bibfnamefont
  {U.}~\bibnamefont {Schneider}}, \ and\ \bibinfo {author} {\bibfnamefont
  {I.}~\bibnamefont {Bloch}},\ }\href {\doibase 10.1126/science.aaa7432}
  {\bibfield  {journal} {\bibinfo  {journal} {Science}\ }\textbf {\bibinfo
  {volume} {349}},\ \bibinfo {pages} {842} (\bibinfo {year}
  {2015})}\BibitemShut {NoStop}%
\bibitem [{\citenamefont {L\"uschen}\ \emph {et~al.}(2017)\citenamefont
  {L\"uschen}, \citenamefont {Bordia}, \citenamefont {Scherg}, \citenamefont
  {Alet}, \citenamefont {Altman}, \citenamefont {Schneider},\ and\
  \citenamefont {Bloch}}]{Luschen17}%
  \BibitemOpen
  \bibfield  {author} {\bibinfo {author} {\bibfnamefont {H.~P.}\ \bibnamefont
  {L\"uschen}}, \bibinfo {author} {\bibfnamefont {P.}~\bibnamefont {Bordia}},
  \bibinfo {author} {\bibfnamefont {S.}~\bibnamefont {Scherg}}, \bibinfo
  {author} {\bibfnamefont {F.}~\bibnamefont {Alet}}, \bibinfo {author}
  {\bibfnamefont {E.}~\bibnamefont {Altman}}, \bibinfo {author} {\bibfnamefont
  {U.}~\bibnamefont {Schneider}}, \ and\ \bibinfo {author} {\bibfnamefont
  {I.}~\bibnamefont {Bloch}},\ }\href {\doibase 10.1103/PhysRevLett.119.260401}
  {\bibfield  {journal} {\bibinfo  {journal} {Phys. Rev. Lett.}\ }\textbf
  {\bibinfo {volume} {119}},\ \bibinfo {pages} {260401} (\bibinfo {year}
  {2017})}\BibitemShut {NoStop}%
\bibitem [{\citenamefont {Sierant}\ \emph {et~al.}(2017)\citenamefont
  {Sierant}, \citenamefont {Delande},\ and\ \citenamefont
  {Zakrzewski}}]{Sierant17}%
  \BibitemOpen
  \bibfield  {author} {\bibinfo {author} {\bibfnamefont {P.}~\bibnamefont
  {Sierant}}, \bibinfo {author} {\bibfnamefont {D.}~\bibnamefont {Delande}}, \
  and\ \bibinfo {author} {\bibfnamefont {J.}~\bibnamefont {Zakrzewski}},\
  }\href {\doibase 10.1103/PhysRevA.95.021601} {\bibfield  {journal} {\bibinfo
  {journal} {Phys. Rev. A}\ }\textbf {\bibinfo {volume} {95}},\ \bibinfo
  {pages} {021601} (\bibinfo {year} {2017})}\BibitemShut {NoStop}%
\bibitem [{\citenamefont {Sierant}\ and\ \citenamefont
  {Zakrzewski}(2018)}]{Sierant18}%
  \BibitemOpen
  \bibfield  {author} {\bibinfo {author} {\bibfnamefont {P.}~\bibnamefont
  {Sierant}}\ and\ \bibinfo {author} {\bibfnamefont {J.}~\bibnamefont
  {Zakrzewski}},\ }\href {\doibase 10.1088/1367-2630/aabb17} {\bibfield
  {journal} {\bibinfo  {journal} {New Journal of Physics}\ }\textbf {\bibinfo
  {volume} {20}},\ \bibinfo {pages} {043032} (\bibinfo {year}
  {2018})}\BibitemShut {NoStop}%
\bibitem [{\citenamefont {Orell}\ \emph {et~al.}(2019)\citenamefont {Orell},
  \citenamefont {Michailidis}, \citenamefont {Serbyn},\ and\ \citenamefont
  {Silveri}}]{Orell19}%
  \BibitemOpen
  \bibfield  {author} {\bibinfo {author} {\bibfnamefont {T.}~\bibnamefont
  {Orell}}, \bibinfo {author} {\bibfnamefont {A.~A.}\ \bibnamefont
  {Michailidis}}, \bibinfo {author} {\bibfnamefont {M.}~\bibnamefont {Serbyn}},
  \ and\ \bibinfo {author} {\bibfnamefont {M.}~\bibnamefont {Silveri}},\ }\href
  {\doibase 10.1103/PhysRevB.100.134504} {\bibfield  {journal} {\bibinfo
  {journal} {Phys. Rev. B}\ }\textbf {\bibinfo {volume} {100}},\ \bibinfo
  {pages} {134504} (\bibinfo {year} {2019})}\BibitemShut {NoStop}%
\bibitem [{\citenamefont {Hopjan}\ and\ \citenamefont
  {Heidrich-Meisner}(2020)}]{Hopjan19}%
  \BibitemOpen
  \bibfield  {author} {\bibinfo {author} {\bibfnamefont {M.}~\bibnamefont
  {Hopjan}}\ and\ \bibinfo {author} {\bibfnamefont {F.}~\bibnamefont
  {Heidrich-Meisner}},\ }\href {\doibase 10.1103/PhysRevA.101.063617}
  {\bibfield  {journal} {\bibinfo  {journal} {Phys. Rev. A}\ }\textbf {\bibinfo
  {volume} {101}},\ \bibinfo {pages} {063617} (\bibinfo {year}
  {2020})}\BibitemShut {NoStop}%
\bibitem [{\citenamefont {Nakamura}\ and\ \citenamefont
  {Lakshmanan}(1986)}]{Nakamura86}%
  \BibitemOpen
  \bibfield  {author} {\bibinfo {author} {\bibfnamefont {K.}~\bibnamefont
  {Nakamura}}\ and\ \bibinfo {author} {\bibfnamefont {M.}~\bibnamefont
  {Lakshmanan}},\ }\href {\doibase 10.1103/PhysRevLett.57.1661} {\bibfield
  {journal} {\bibinfo  {journal} {Phys. Rev. Lett.}\ }\textbf {\bibinfo
  {volume} {57}},\ \bibinfo {pages} {1661} (\bibinfo {year}
  {1986})}\BibitemShut {NoStop}%
\end{thebibliography}
%

\end{document}